\documentclass[aps,prd,twocolumn,showpacs,eqsecnum,amsmath,nofootinbib]{revtex4}
\usepackage{graphicx}

\begin{document}
%\draft
\def \d {{\rm d}}
\def \p {{\rm p}}
\def \im {{\rm i}}
\def \A {{\rm A}}
\def \B {{\rm B}}
\def \Z {{\rm Z}}
\def \U {{\cal U}}
\def \V {{\cal V}}
\newcommand{\pul}{{\textstyle{\frac{1}{2}}}}      % 1/2  mensi pismo
\newcommand{\dpul}{{\textstyle{\frac{1}{2}\delta}}}   % delta/2
\newcommand{\ddctvrt}{{\textstyle{\frac{1}{4}\delta^2}}}  % delta^2/4
\newcommand{\rovno}{\!\!\!\!& = &\!\!\!\!}              % rovnitko se zarovnanim pro pouztiti v eqnarray

\title{Refraction of geodesics by impulsive spherical gravitational waves in constant-curvature spacetimes with a cosmological constant}

\author{Ji\v{r}\'{\i} Podolsk\'y}
\email{podolsky@mbox.troja.mff.cuni.cz}
\author{Robert \v{S}varc}
\email{robert.svarc@mff.cuni.cz}
\affiliation{%%
  Institute of Theoretical Physics,  Faculty of Mathematics and Physics, Charles University in Prague,\\
  V Hole\v{s}ovi\v{c}k\'{a}ch 2, 180 00 Prague 8, Czech Republic
  }

\date{\today}

\begin{abstract}
We investigate motion of test particles in exact spacetimes with an expanding impulsive gravitational wave which propagates
in Minkowski, de~Sitter or anti-de~Sitter universe. Using the continuous form of these metrics we derive explicit junction conditions
and simple refraction formulae for null, timelike and spacelike geodesics crossing a general impulse of this type.
In particular, we present a detailed geometrical description of the motion of test particles in a special class of axially
symmetric spacetimes in which the impulse is generated by a snapped cosmic string.

\end{abstract}


% \keywords{}

\maketitle

\section{Introduction}\label{sec:intro}

In the fundamental work \cite{Pen72} Roger Penrose introduced an elegant geometric ``cut and paste'' method for construction of impulsive spherical gravitational waves in a flat background. This is based on cutting Minkowski space along a null cone and then re-attaching the two
pieces with a suitable warp. An explicit class of such spacetimes, using coordinates in which the metric functions are continuous across the impulse, was subsequently given by Nutku and Penrose \cite{NutPen92}, Hogan \cite{Hogan93,Hogan94} and, to include a nonvanishing cosmological constant, in \cite{Hogan92,PodGri99,PodGri00}. An additional acceleration parameter can also be introduced \cite{AliNut01}.

This gives the complete family of expanding spherical waves of a very short duration which propagate in Minkowski, de Sitter or anti-de~Sitter universe, that is in spacetimes with a constant curvature (zero, positive or negative, respectively). Such solutions can naturally be understood as impulsive limits of Robinson--Trautman type-N vacuum solutions \cite{Stephanietal:book, GriffithsPodolsky:book}, namely a suitable family of spherical sandwich waves of this type \cite{PodGri99,GriPodDoc02}.

A stereographic interpretation of complex spatial coordinate involved in the Penrose junction condition across the impulse can be used for an explicit construction of specific solutions of this type, in particular those which describe impulsive spherical waves generated by colliding and snapping cosmic strings \cite{PodGri00}. First such solution given already in \cite{NutPen92} represents the snapping of a cosmic string, identified by a deficit angle in the region outside the spherical impulsive gravitational wave. The collision and breaking of a pair of cosmic strings can also be described in this way.

The particular solution for a spherical gravitational impulse generated by a snapping cosmic string in Minkowski space was alternatively described by Bi\v{c}\'ak and Schmidt~\cite{BicSch89}. This was obtained as a limiting case of the Bonnor--Swaminarayan solution for an infinite acceleration of a pair of Curzon--Chazy particles (see Chapter~15 of \cite{GriffithsPodolsky:book}). It was observed in \cite{Bicak90} that such situation is equivalent to the splitting of an infinite cosmic string as described in~\cite{GlePul89} or, rather, of two semi-infinite cosmic strings approaching at the speed of light and separating again at the instant at which they ``collide''.

The same explicit solution was also obtained in the limit of an infinite acceleration in the more general class which represents a pair of uniformly accelerating particles with an arbitrary multipole structure \cite{PodGri01a}, or as an analogous limit of the C-metric which describes accelerating black holes \cite{PodGri01b}. In the latter case, a nonvanishing cosmological constant can also be considered. This leads to a specific expanding spherical impulse generated by a snapping cosmic string in the (anti-)de~Sitter universe \cite{PodGri04}.

More details concerning these impulsive metrics and other references can be found in the review works \cite{Pod02b,BarHog03a} and in Chapter~20 of \cite{GriffithsPodolsky:book}. Note also that particle creation and other quantum effects in such spacetimes were investigated, e.~g., by Horta\c csu and his collaborators \cite{Hort90,Hort95,Hort96a,Hort96b}.

The main objective of the present work is to study specific properties of these spacetimes, namely the motion of test particles influenced by the spherical impulsive waves. In fact, Podolsk\'y and Steinbauer in \cite{PodSte03} already investigated and described the behavior of exact geodesics in the case when the impulse expands in Minkowski flat space. Here we will generalize this study to any value of the cosmological constant, i.e., we will analyze the effects on geodesics when the spherical impulse expands in de Sitter or anti-de Sitter universe.
Moreover, we will present the results in a form which is more convenient for physical and geometric interpretation.

Our paper is organized as follows. In Sec.~\ref{sec:eiw} we review the class of spacetimes under consideration and describe the geometry of the expanding impulses.
By employing a continuous form of the metric, in Sec.~\ref{sec:geos} we investigate
a large class of~$C^1$ geodesics crossing the spherical impulse. We explicitly derive the junction conditions and the
refraction formulae, we study a subfamily of privileged global geodesics and rewrite
the junction conditions in a convenient five-dimensional formalism when ${\Lambda\not=0}$.
In Sec.~\ref{sec:string} we focus on impulsive waves generated by a snapped
cosmic string.  We discuss in detail the physical and geometric interpretation of the motion of test
particles influenced by an impulse of such type.

\section{Expanding impulsive waves in constant-curvature backgrounds}
\label{sec:eiw}

As a natural background for constructing the family of spherical expanding impulsive waves we consider the conformally flat metric
\begin{equation}
   \d s_0^2= \frac{ 2\,\d\eta\,\d\bar\eta-2\,\d\U\,\d\V  }
   {[\,1+\frac1{6}\Lambda(\eta\bar\eta-\U\V)\,]^2} \, .
 \label{confunif}
\end{equation}
This is a unified form for all spaces of constant curvature, namely Minkowski space when ${\Lambda=0}$, de~Sitter space when ${\Lambda>0}$, and anti-de~Sitter space when ${\Lambda<0}$.

Indeed, with the standard representation of the double null coordinates
\begin{equation}
   \U=\textstyle{\frac{1}{\sqrt2}}(t-z),\
   \V=\textstyle{\frac{1}{\sqrt2}}(t+z),\
   \eta=\textstyle{\frac{1}{\sqrt2}}(x+\,\im\,y),
\label{trmink2}
\end{equation}
the metric~\eqref{confunif} reads
 \begin{equation}
   \d s_0^2 = \frac{-\d t^2+\d x^2+\d y^2+\d z^2}
   {[\,1+\frac1{12}\Lambda(-t^2+x^2+y^2+z^2)\,]^2}\,,
 \label{mink}
 \end{equation}
which for ${\Lambda=0}$ is the familiar form of the flat space. In the case ${\Lambda\not=0}$, it is well-known that the corresponding de~Sitter and  anti-de~Sitter spaces can be represented as a four-dimensional hyperboloid
 \begin{equation}
   -\Z_0^2 +\Z_1^2 +\Z_2^2 +\Z_3^2 +\varepsilon\Z_4^2 =\varepsilon a^2,
 \label{hyperboloid}
 \end{equation}
embedded in a flat five-dimensional spacetime
 \begin{equation}
   \d s_0^2= -\d\Z_0^2 +\d\Z_1^2 +\d\Z_2^2 +\d\Z_3^2 +\varepsilon\d\Z_4^2\,,
\label{5Dflatspace}
 \end{equation}
where ${\varepsilon=1}$ for the de~Sitter space (${\Lambda>0}$), ${\varepsilon=-1}$ for the anti-de~Sitter space (${\Lambda<0}$), and ${a=\sqrt{3/|\Lambda|}}$. The specific parametrisation of \eqref{hyperboloid} given as
\begin{eqnarray}
   && \Z_0 = {\textstyle\frac1{\sqrt2}(\V+\U)\left[1+\frac1{6}\Lambda(\eta\bar\eta-\U\V)\right]^{-1}}, \nonumber\\
   && \Z_1 = {\textstyle\frac1{\sqrt2}(\V-\U)\left[1+\frac1{6}\Lambda(\eta\bar\eta-\U\V)\right]^{-1}}, \nonumber\\
   && \Z_2 = {\textstyle\frac1{\sqrt2}(\eta+\bar\eta)\left[1+\frac1{6}\Lambda(\eta\bar\eta-\U\V)\right]^{-1}} , \label{ZcoordsCX}\\
   && \Z_3 = {\textstyle \frac{-\im}{\sqrt2}(\eta-\bar\eta)\left[1+\frac1{6}\Lambda(\eta\bar\eta-\U\V)\right]^{-1}} , \nonumber\\
   && \Z_4 = a{\textstyle \left[1-\frac1{6}\Lambda(\eta\bar\eta-\U\V)\right]\!\!\left[1+\frac1{6}\Lambda(\eta\bar\eta-\U\V)\right]^{-1}}, \nonumber
\end{eqnarray}
or inversely
\begin{eqnarray}
   && \U={\sqrt2\,a}\frac{\Z_0-\Z_1}{\Z_4+a}\,,\nonumber\\
   && \V={\sqrt2\,a}\frac{\Z_0+\Z_1}{\Z_4+a}\,,\label{invZcoords}\\
   && \eta={\sqrt2\, a}\frac{\Z_2+\im\,\Z_3}{\Z_4+a}\,,\nonumber
\end{eqnarray}
takes \eqref{5Dflatspace} to the metric form (\ref{confunif}). Consequently, for ${\U,\V\in(-\infty,+\infty)}$ and $\eta$ an arbitrary complex number, these coordinates cover the entire (anti-)de~Sitter manifold (except the coordinate singularities at ${\U,\V=\infty}$). 
For more details about these coordinates and other properties of maximally symmetric spacetimes see Chapters~3--5 of~\cite{GriffithsPodolsky:book}.

The Penrose ``cut and paste'' method \cite{Pen72} for constructing impulsive spherical waves in such backgrounds of constant curvature can now be performed explicitly as follows (see \cite{Hogan92,PodGri00}).

In the region ${U \ge 0}$, let us consider the transformation
 \begin{eqnarray}
 \V   &=& \V^+ = AV-DU\,,   \nonumber\\
 \U   &=& \U^+ = BV-EU\,,   \label{transe3} \\
 \eta &=& \eta^+\,= CV-FU\,,  \nonumber
 \end{eqnarray}
to coordinates ${(U,V,Z, \bar Z)}$, where
 \begin{eqnarray}
&&A= \frac{1}{p|h'|}\,,\qquad
B= \frac{|h|^2}{p|h'|}\,,\qquad
C= \frac{h}{ p|h'|}\,,   \nonumber\\
&&D= \frac{1}{|h'|}\left\{
\frac{p}{4} \left|\frac{h''}{h'}\right|^2+\epsilon
\left[1+\frac{Z}{2}\frac{h''}{h'}+\frac{\bar Z}{2}\frac{\bar h''}{\bar h'}
\right]\right\},\nonumber\\
&&E= \frac{|h|^2}{|h'|}
\bigg\{ \frac{p}{4}\left|\frac{h''}{h'}-2\frac{h'}{h}\right|^2  \label{transe4}\\
&&\qquad+\epsilon\left[ 1+\frac{Z}{2}
\left(\frac{h''}{h'}-2\frac{h'}{h}\right)+\frac{\bar Z}{2}
\left(\frac{\bar h''}{\bar h'}-2\frac{\bar h'}{\bar h}\right)
\right]\bigg\}, \nonumber\\
&&F= \frac{h}{|h'|}\bigg\{
\frac{p}{4}\left(\frac{h''}{h'}-2\frac{h'}{h}\right)
\frac{\bar h''}{\bar h'}  \nonumber\\
&&\qquad+\epsilon\left[1+
 \frac{Z}{2}\left(\frac{h''}{h'}-2\frac{h'}{h}\right)
+\frac{\bar Z}{2}\frac{\bar h''}{\bar h'}\right]\bigg\},
\nonumber
 \end{eqnarray}
with
\begin{equation}
   p=1+\epsilon Z\bar Z\,, \qquad \epsilon=-1,0,+1
   \label{pandep}
\end{equation}
(the parameter $\epsilon$ is the Gaussian curvature of the spatial 2-surfaces in the closely related Robinson--Trautman foliation of the spacetimes, cf.~Sec.~19.2 of~\cite{GriffithsPodolsky:book}). Here
\begin{equation}
   h \equiv h(Z)\,,
   \label{hofZ}
\end{equation}
is an \emph{arbitrary} complex function, and the derivative with respect to its argument $Z$ is denoted by a prime. The Minkowski and (anti-)de~Sitter metric (\ref{confunif}) then becomes
 \begin{equation}
 \d s_0^2 = \frac{2\left| (V/p)\,\d Z+U p\,\bar H\,\d\bar Z \right|^2 +2\,\d U\,\d V -2\epsilon\,\d U^2}
   {[\,1+\frac1{6}\Lambda \,U( V-\epsilon U)\,]^2}\,,
 \label{U>0}
 \end{equation}
 where $H$ is the Schwarzian derivative of $h$ given as
 \begin{equation}
 H(Z)=\frac{1}{2} \left[\frac{h'''}{h'}-\frac{3}{2}\left(\frac{h''}{h'}\right)^2
 \right] .
 \label{Schwarz}
 \end{equation}

In the complementary region ${U \le 0}$, we apply a highly simplified form of the transformation \eqref{transe3} which arises for the special choice of the function ${h(Z)=Z}$. In view of \eqref{transe4}, this implies relations
\begin{eqnarray}
   \V   &=& \V^- = \frac{V}{p}-\epsilon U \,,   \nonumber\\
   \U   &=& \U^- = \frac{|Z|^2}{p}\,V-U\,,   \label{inv} \\
   \eta &=& \eta^-\,= \frac{Z}{p}\,V\,.  \nonumber
\end{eqnarray}
Since ${H=0}$ in this case, by applying the transformation (\ref{inv}) the metric (\ref{confunif}) takes the form
 \begin{equation}
 \d s_0^2 = \frac{2\, (V/p)^2\,\d Z\, \d \bar Z+ 2\,\d U\,\d V -2\epsilon\,\d U^2}
   {[\,1+\frac1{6}\Lambda \,U( V-\epsilon U)\,]^2}\,.
 \label{U<0}
 \end{equation}

\begin{figure}[ht]
\begin{center}
\includegraphics[scale=0.51]{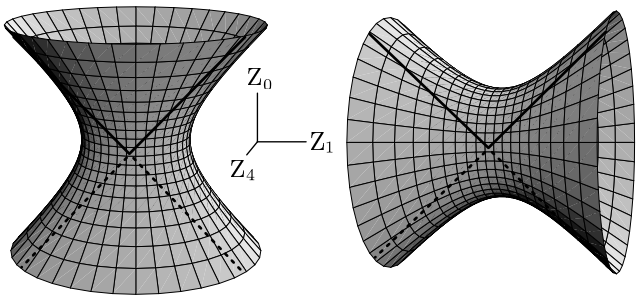}%\
%\vspace{55mm}
\end{center}
\caption{\label{figure1}%
An expanding spherical impulse can be visualized as a section  ${\Z_4=a}$ of the four-dimensional hyperboloids representing de~Sitter (left)
and anti-de~Sitter (right) spaces. The bold lines are trajectories of opposite poles of an expanding spherical wave surface given by ${\Z_2=0=\Z_3}$.
The time-reversed situation in the region ${\Z_0<0}$, indicated by dashed lines, corresponds to contracting impulsive waves. }
\end{figure}

Both in the coordinates of (\ref{U>0}) and in the ones used in (\ref{U<0}), the boundary hypersurface ${U=0}$ is a \emph{null cone} given by
${\,\eta\bar\eta-\U\V=0}$. Using \eqref{trmink2}, it is obviously an expanding sphere ${x^2+y^2+z^2=t^2}$ in flat Minkowski space.
In view of \eqref{invZcoords}, it is also an expanding sphere ${\Z_1^2 +\Z_2^2 +\Z_3^2=\Z_0^2}$ in the (anti-)de~Sitter universe.
Considering the relation \eqref{hyperboloid} it follows that such null hypersurface ${U=0}$ is the vertical cut ${\Z_4=a}$ through the
de~Sitter and anti-de~Sitter hyperboloid in a flat five-dimensional spacetime, as shown in Fig.~\ref{figure1}.
This represents a spherical impulse which originates at time ${\Z_0=0}$ and subsequently for ${\Z_0>0}$ expands with the speed of light
in these backgrounds (alternatively, for ${\Z_0<0}$ the impulse is contracting).

\vspace{5mm}

An explicit \emph{global} metric which is continuous across the impulse at ${U=0}$ is now easily obtained by attaching the line element (\ref{U<0}) for $U<0$ to (\ref{U>0}) for $U>0$.
The resulting metric takes the form
\begin{equation}
   \d s^2 \!=\! \frac{2\!\left| (V/p)\d Z+ U\Theta(U) p\bar H\d\bar Z \right|^2 +2\,\d U\d V -2\epsilon\,\d U^2}
   {[\,1+\frac1{6}\Lambda \,U( V-\epsilon U)\,]^2}\,,
  \label{en0}
 \end{equation}
where ${\Theta(U)}$ is the Heaviside step function. Such combined metric is continuous, but the discontinuity in the derivatives of the metric functions across ${U=0}$ yields an impulsive gravitational wave term in the curvature proportional to the Dirac $\delta$-distribution. More precisely, in a suitable null tetrad, the only nonvanishing component of the Weyl tensor is ${\Psi_4=(p^2H/ V)\,\delta(U)}$ (for more details see~\cite{PodGri00}). The spacetime is thus conformally flat everywhere except on the impulsive wave surface ${U=0}$. Also, the only nonvanishing tetrad component of the Ricci tensor is ${\Phi_{22}=(p^4H\bar H/ V^2)\,U\delta(U)}$. This demonstrates that the spacetime is vacuum everywhere, except on the impulse at ${V=0}$ and at possible singularities of the function $p^2H$.

The expanding spherical impulse located at ${U=0}$ obviously splits the spacetime into two separate conformally flat vacuum regions
(Minkowski, de~Sitter, or anti-de~Sitter, according to ${\Lambda}$). For brevity, in the following we shall denote the constant-curvature half-space
${U>0}$ as being \emph{``in front of the wave''}, and the other constant-curvature half-space ${U<0}$  as being \emph{``behind the wave''}.

\section{Geodesic motion in spacetimes with expanding impulsive waves}
\label{sec:geos}

The purpose of this paper is to investigate the effect of expanding impulsive waves on motion of freely moving test particles.
We start by recalling geodesics in Minkowski and (anti-)de~Sitter spaces, then we will derive junction conditions for complete 
geodesics in the impulsive spacetimes summarized in the previous section and we will present the refraction formulae.

\subsection{Geodesics in the backgrounds}
\label{subsec:geosbackg}

Geodesic motion in spaces of constant curvature~\eqref{mink}, the background spaces in which an impulse propagates, is well known.

When ${\Lambda=0}$, this is just flat Minkowski space. General geodesics are, of course,
given by
\begin{eqnarray}
&& t = \gamma\,\tau\,,  \nonumber\\
&& x = x_i+\dot x_i\,(\tau-\tau_i)\,, \nonumber\\
&& y = y_i+\dot y_i\,(\tau-\tau_i)\,, \label{geod+}\\
&& z = z_i+\dot z_i\,(\tau-\tau_i)\,, \nonumber
\end{eqnarray}
with ${\gamma=\sqrt{\dot x_i^2+\dot y_i^2+\dot z_i^2-e}}$, i.e., $\tau$ is a normalized affine parameter of timelike (${e=-1}$) or spacelike (${e=+1}$) geodesics. For null geodesics (${e=0}$) it is always possible to scale the factor $\gamma$ to unity.  The constants ${x_i,y_i,z_i}$ and ${\dot x_i,\dot y_i, \dot z_i}$ characterize the position and velocity, respectively, of each test particle at the instant
\begin{equation}
\tau_i=\frac{1}{\gamma}\,\sqrt{x_i^2+y_i^2+z_i^2}\,, \label{taui}
\end{equation}
when the geodesic intersects the null cone ${U=0}$. At $\tau_i$ each particle is hit by the impulse and its trajectory is refracted, see Sec.~\ref{subsec:refrac}.

In the case of a nonvanishing cosmological constant~$\Lambda$, to express all geodesics in the corresponding de~Sitter and anti-de~Sitter spaces
it is very useful to employ the five-dimensional formalism. It can be shown~\cite{PodOrt01} that, using the coordinates of~\eqref{5Dflatspace},
the explicit geodesic equations have a very simple and unified form, namely ${\ddot \Z_\p+ \frac{1}{3}\Lambda\,e\,\Z_\p=0}$, where  ${\p=0,1,2,3,4}$.
Thus, explicit geodesics on the hyperboloid~\eqref{hyperboloid} are
\begin{eqnarray}
&&  \hskip-13.5mm  \Z_\p = \Z_{\p i}+\dot \Z_{\p i}\,(\tau-\tau_i)
  \hskip16.5mm\hbox{ when}\   \varepsilon e=0\,, \label{g21}\\
&&  \hskip-13.5mm  \Z_\p = \Z_{\p i}\cosh\Big(\frac{\tau-\tau_i}{a}\Big)+a\dot \Z_{\p i}\sinh\Big(\frac{\tau-\tau_i}{a}\Big)\label{g22}\\
&&  \hskip40mm  \hbox{when}\ \varepsilon e<0\,, \nonumber\\
&&  \hskip-13.5mm  \Z_\p = \Z_{\p i}\cos\Big(\frac{\tau-\tau_i}{a}\Big)+a\dot \Z_{\p i}\sin\Big(\frac{\tau-\tau_i}{a}\Big)\label{g23}\\
&&  \hskip40mm \hbox{when}\ \varepsilon e>0\,, \nonumber
\end{eqnarray}
where ${a=\sqrt{3/|\Lambda|}}$. The relation~\eqref{g21} describes null geodesics, expression~\eqref{g22} represents timelike geodesics
in de~Sitter space (${\varepsilon=1}$) or spacelike geodesics in anti-de~Sitter space (${\varepsilon=-1}$), whereas~\eqref{g23} corresponds
to spacelike/timelike geodesics in de~Sitter/anti-de~Sitter space, respectively. Here  $\tau$ is an affine parameter and $\Z_{\p i}$, $\dot \Z_{\p i}$
are constants of integration, namely the positions and velocities at the instant of interaction with the impulse ${\tau=\tau_i}$.
These ten constants are constrained by the following three conditions
\begin{eqnarray}
&&  \hskip-13.5mm  -(\dot{\Z}_{0i})^2+(\dot{\Z}_{1i})^2+(\dot{\Z}_{2i})^2+(\dot{\Z}_{3i})^2+\varepsilon(\dot{\Z}_{4i})^2 = e
 \,, \label{con1}\\
&&  \hskip-13.5mm  -(\Z_{0i})^2+(\Z_{1i})^2+(\Z_{2i})^2+(\Z_{3i})^2+\varepsilon(\Z_{4i})^2 = \varepsilon a^2
  , \label{con2}\\
&&  \hskip-13.5mm  - \Z_{0i}\dot{\Z}_{0i}+\Z_{1i}\dot{\Z}_{1i}+\Z_{2i}\dot{\Z}_{2i}+\Z_{3i}\dot{\Z}_{3i}+\varepsilon \Z_{4i}\dot{Z}_{4i} = 0 \,.
 \label{con3}
\end{eqnarray}
Eq.~\eqref{con1} is the normalization of the affine parameter, Eq.~\eqref{con2} follows from the constraint \eqref{hyperboloid}, and Eq.~\eqref{con3} from its derivative.

By combining relations~\eqref{invZcoords} and \eqref{g21}--\eqref{g23} it is now straightforward to express explicitly all geodesics in the four-dimensional metric representation of (anti-)de~Sitter universe~\eqref{confunif}. Considering~\eqref{trmink2}, which implies
\begin{equation}
t = \frac{2a\,\Z_0}{\Z_4+a}\,,\   z = \frac{2a\,\Z_1}{\Z_4+a}\,,\ x = \frac{2a\, \Z_2}{\Z_4+a}\,,\   y =\frac{ 2a\, \Z_3}{\Z_4+a}\,,
   \label{realZcoords}
\end{equation}
 we also obtain geodesics in the metric~\eqref{mink}, and by using other parametrisations of the hyperboloid~\eqref{hyperboloid}, as summarised in~\cite{GriffithsPodolsky:book}, we may easily derive geodesics in any standard metric form of these constant-curvature spacetimes. Some of them will be given below.

Notice finally that close to the impulse (where ${\tau-\tau_i}$ is small) and also in the limit ${\Lambda\to0}$ (so that ${1/a}$ is small) expressions
\eqref{g21}--\eqref{g23} take the same linear form ${\Z_\p \approx \Z_{\p i}+\dot \Z_{\p i}\,(\tau-\tau_i)}$. In view of \eqref{realZcoords} this is
fully consistent with Eqs. \eqref{geod+}.

\subsection{Explicit continuation of geodesics across the impulse}
\label{subsec:junction}

Now we will investigate geodesics in complete spacetimes \eqref{en0} with the  wave localized on ${U=0}$. Geodesics which pass through the impulse
have the same form \eqref{geod+} or \eqref{g21}--\eqref{g23} both in front of the impulse and behind it. However, the constants of integration
${\Z_{\p i}, \dot \Z_{\p i}}$ may have \emph{different values} on both sides.

We thus have to find explicit relations between these constants. To apply the appropriate junction conditions, we \emph{assume that the
geodesics are  $C^1$ across the impulse} in the continuous coordinate system of \eqref{en0}. It means that the corresponding functions ${Z(\tau),V(\tau),U(\tau)}$
and also their first derivatives with respect to the affine parameter $\tau$, evaluated at the interaction time $\tau=\tau_i$ (such that ${U(\tau_i)=0}$), are
continuous across the impulse. With this assumption, the constants
\begin{eqnarray}
&&Z_i\equiv Z(\tau_i)\ ,\quad V_i\equiv V(\tau_i)\,,\quad  U_i\equiv U(\tau_i)=0\,,\nonumber\\
&&\dot Z_i\equiv \dot Z(\tau_i)\,,\quad\,\dot V_i\equiv \dot V(\tau_i)\,,\quad \dot U_i\equiv \dot U(\tau_i)\,,
\label{const}
\end{eqnarray}
describing positions and velocities at $\tau_i$ have the \emph{same} values when evaluated in the limits $U\to0$ both from the region in front (${U>0}$)
and behind the impulse (${U<0}$).

To express the corresponding values in the conformally flat coordinates of \eqref{confunif}, it is now straightforward to substitute \eqref{const} into
the transformations \eqref{transe3} and \eqref{inv},
\begin{eqnarray}
\V_i^+ \!&=&\! AV_i \,, \qquad\quad \V_i^-= \frac{V_i}{p} \,, \nonumber \\
\U_i^+ \!&=&\! BV_i \,, \qquad\quad \U_i^-= \frac{|Z_i|^2}{p}V_i \,, \label{C1 polohy pred/za vlnou} \\
\eta_i^+\!&=&\! CV_i \,, \qquad\quad \eta_i^-= \frac{Z_i}{p}V_i \,, \nonumber
\end{eqnarray}
and their derivatives,
\begin{eqnarray}
\dot{\V}_i^+ &=& V_i(A_{,Z}\dot{Z}_i+A_{,\bar{Z}}\dot{\bar{Z}}_i)+A\dot{V}_i-D\dot{U}_i \,, \nonumber \\
\dot{\U}_i^+ &=& V_i(B_{,Z}\dot{Z}_i+B_{,\bar{Z}}\dot{\bar{Z}}_i)+B\dot{V}_i-E\dot{U}_i \,, \label{C1 rychlosti pred vlnou} \\
\dot{\eta}_i^+ &=& V_i(C_{,Z}\dot{Z}_i+C_{,\bar{Z}}\dot{\bar{Z}}_i)+C\dot{V}_i-F\dot{U}_i \,,  \nonumber
\end{eqnarray}
\begin{eqnarray}
\dot{\V}_i^- &=& -\frac{\epsilon V_i}{p^2}(Z_i\dot{\bar{Z}}_i+\bar{Z}_i\dot{Z}_i)+\frac{\dot{V}_i}{p}-\epsilon\dot{U}_i \,, \nonumber \\
\dot{\U}_i^- &=& \frac{V_i}{p^2}(Z_i\dot{\bar{Z}}_i+\bar{Z}_i\dot{Z}_i)+\frac{|Z_i|^2}{p}\dot{V}_i-\dot{U}_i \,, \label{C1 rychlosti za vlnou} \\
\dot{\eta}_i^-&=& \frac{V_i}{p^2}(\dot{Z}_i-\epsilon Z_i Z_i\dot{\bar{Z}}_i)+\frac{Z_i}{p}\dot{V}_i \,, \nonumber
\end{eqnarray}
respectively (here $A, B, C, D, E, F, p$ and their derivatives are constants, namely the coefficients \eqref{transe4}, \eqref{pandep} evaluated at ${Z=Z_i}$).

Since we wish to express the ``$-$'' parameters behind the impulse in terms of the ``$+$'' parameters in front of the impulse, we invert
expressions \eqref{C1 polohy pred/za vlnou}, \eqref{C1 rychlosti pred vlnou} in the half-space in front of the wave, which yields
\begin{eqnarray}
&& h(Z_i) = \frac{\eta_i^+}{\V_i^+} \,, \label{spojite interakcni parametry pomoci parametru pred vlnou}\\
&& V_i = \frac{\U_i^+}{B}=\frac{\V_i^+}{A}=\frac{\eta_i^+}{C} \,, \nonumber \\
&& \dot{Z}_i = \frac{p^2}{V_i}\left(\bar{C}_{,\bar{Z}}\,\dot{\eta}_i^+
    +C_{,\bar{Z}}\,\dot{\bar{\eta}}_i^+-A_{,\bar{Z}}\,\dot{\U}_i^+-B_{,\bar{Z}}\,\dot{\V}_i^+\right), \nonumber \\
&& \dot{V}_i = D\,\dot{\U}_i^++E\,\dot{\V}_i^+-\bar{F}\,\dot{\eta}_i^+-F\,\dot{\bar{\eta}}_i^++2\epsilon\,\dot{U}_i \,, \nonumber \\
&&\dot{U}_i = \frac{1}{V_i}\left(\bar{\eta}_i^+\dot{\eta}_i^++\eta_i^+\dot{\bar{\eta}}_i^+-\V_i^+\dot{\U}_i^+-\U_i^+\dot{\V}_i^+\right). \nonumber
\end{eqnarray}
In order to obtain these relations, we employed the identities valid for the coefficients \eqref{transe4},
\begin{eqnarray}
AB-C\bar{C}&=& 0 , \nonumber \\
DE-F\bar{F} &=& \epsilon , \\
AE+BD-C\bar{F}-\bar{C}F &=& 1 , \nonumber
\end{eqnarray}
and also for their derivatives
\begin{eqnarray}
DE_{,Z}+D_{,Z}E-F\bar{F}_{,Z}-F_{,Z}\bar{F} &=& 0 , \nonumber \\
A_{,Z}E+B_{,Z}D-C_{,Z}\bar{F}-\bar{C}_{,Z}F &=& 0 , \nonumber \\
A\,E_{,Z}+BD_{,Z}-C\bar{F}_{,Z}-\bar{C}F_{,Z} &=& 0 , \nonumber \\
A_{,Z}B+AB_{,Z}-C_{,Z}\bar{C}-C\bar{C}_{,Z} &=& 0 , \nonumber \\
A_{,Z}B_{,Z}-C_{,Z}\bar{C}_{,Z} &=& 0 , \nonumber \\
D_{,Z}E_{,Z}-F_{,Z}\bar{F}_{,Z} &=& 0 , \\
A_{,Z}E_{,Z}+B_{,Z}D_{,Z}-C_{,Z}\bar{F}_{,Z}-\bar{C}_{,Z}F_{,Z} &=& H , \nonumber \\
C_{,Z}\bar{C}_{,\bar{Z}}+C_{,\bar{Z}}\bar{C}_{,Z}-A_{,Z}B_{,\bar{Z}}-A_{,\bar{Z}}B_{,Z} &=& \frac{1}{p^2} , \nonumber \\
F_{,Z}\bar{F}_{,\bar{Z}}+F_{,\bar{Z}}\bar{F}_{,Z}-D_{,Z}E_{,\bar{Z}}-D_{,\bar{Z}}E_{,Z} &=& p^2|H|^2 ,\nonumber \\
A_{,Z}E_{,\bar{Z}}+A_{,\bar{Z}}E_{,Z}+B_{,Z}D_{,\bar{Z}}+B_{,\bar{Z}}D_{,Z}\qquad&& \nonumber \\
-C_{,Z}\bar{F}_{,\bar{Z}}-C_{,\bar{Z}}\bar{F}_{,Z}-\bar{C}_{,Z}F_{,\bar{Z}}-\bar{C}_{,\bar{Z}}F_{,Z}&=& 0 ,\nonumber
\end{eqnarray}
plus their complex conjugates.

Now it only remains to substitute \eqref{spojite interakcni parametry pomoci parametru pred vlnou} into the expressions for positions \eqref{C1 polohy pred/za vlnou} and velocities
(\ref{C1 rychlosti za vlnou}) behind the impulse. For the positions we thus obtain
\begin{eqnarray}
&& \V_i^-=|h'|\,\V_i^+ , \nonumber\\
&& \U_i^-=|h'|\frac{|Z_i|^2}{|h|^2}\,\U_i^+ , \label{C1 polohy za vlnou pomoci parametru pred vlnou}\\
&& \eta_i^-=|h'|\frac{Z_i}{h}\,\eta_i^+ , \nonumber
\end{eqnarray}
while for the velocities, after straightforward but somewhat lengthy calculation, we get
\begin{eqnarray}
&& \dot{\V}_i^- = b_{_\V}\dot{\V}_i^+ +a_{_\V}\dot{\U}_i^+ +\bar{c}_{_\V}\dot{\eta}_i^+ +c_{_\V}\dot{\bar{\eta}}_i^+ , \nonumber\\
&& \dot{\U}_i^- = b_{_\U}\dot{\V}_i^+ +a_{_\U}\dot{\U}_i^+ +\bar{c}_{_\U}\dot{\eta}_i^+ +c_{_\U}\dot{\bar{\eta}}_i^+ , \label{C1 rychlosti za vlnou pomoci parametru pred vlnou - preznacene} \\
&& \dot{\eta}_i^- \,= b_{\eta} \dot{\V}_i^+ +a_{\eta}\,\dot{\U}_i^+ +\bar{c}_{\eta}\,\dot{\eta}_i^+ +c_{\eta}\,\dot{\bar{\eta}}_i^+ , \nonumber
\end{eqnarray}
where
\begin{eqnarray}
&& b_{_\V} = \frac{|h|^2}{4|h'|}\left|\frac{h''}{h'}-2\frac{h'}{h}\right|^2,  \nonumber \\
&& a_{_\V} = \frac{1}{4|h'|}\left|\frac{h''}{h'}\right|^2 , \label{acalV}\\
&& c_{_\V} = -\frac{h}{4|h'|}\left(\frac{h''}{h'}-2\frac{h'}{h}\right)\frac{\bar{h}''}{\bar{h}'} ,   \nonumber
\end{eqnarray}
\begin{eqnarray}
&& b_{_\U} =  \frac{|h|^2}{|h'|}\left|1+\frac{Z_i}{2}\bigg(\frac{h''}{h'}-2\frac{h'}{h}\bigg) \right|^2, \nonumber\\
&& a_{_\U} = \frac{1}{|h'|}\left|1+\frac{Z_i}{2}\frac{h''}{h'}\right|^2,  \label{ccalU} \\
&& c_{_\U} = -\frac{h}{|h'|}\bigg[1+\frac{Z_i}{2}\left(\frac{h''}{h'}-2\frac{h'}{h}\right)\bigg]\bigg[1+\frac{\bar{Z}_i}{2}\frac{\bar{h}''}{\bar{h}'}\bigg],\nonumber  
\end{eqnarray}
\begin{eqnarray}
&& b_{\eta} = \frac{|h|^2}{2|h'|}\bigg[1+\frac{Z_i}{2}\bigg(\frac{h''}{h'}-2\frac{h'}{h}\bigg)\bigg]\bigg(\frac{\bar{h}''}{\bar{h}'}-2\frac{\bar{h}'}{\bar{h}}\bigg),\nonumber\\
&& a_{\eta}= \frac{1}{2|h'|}\bigg(1 +\frac{Z_i}{2}\frac{h''}{h'}\bigg)\frac{\bar{h}''}{\bar{h}'}, \nonumber\\
&& \bar{c}_{\eta}= -\frac{\bar{h}}{2|h'|}\bigg(1+\frac{Z_i}{2} \frac{h''}{h'}\bigg)\bigg(\frac{\bar{h}''}{\bar{h}'}-2\frac{\bar{h}'}{\bar{h}}\bigg), \nonumber\\
&& c_{\eta}= -\frac{h}{2|h'|}\bigg[1+\frac{Z_i}{2}\left(\frac{h''}{h'}-2\frac{h'}{h}\right) \bigg]\frac{\bar{h}''}{\bar{h}'} , \label{ccalbareta} 
\end{eqnarray}
and, naturally, ${\bar{c}_{_\V} = \overline{c_{_\V}}}$,  ${\bar{c}_{_\U} = \overline{c_{_\U}}}$. Again, all these coefficients
are constants which are obtained by evaluating the function $h$ and its derivatives at ${Z=Z_i}$. Interestingly, they do not depend on the
parameter $\epsilon$. For ${h(Z)=Z}$ there is no refraction effect, which is consistent with the fact that ${H=0}$, i.e., the impulse is absent.

Finally, using the transformation \eqref{trmink2} we may rewrite the expressions for junction conditions \eqref{C1 polohy za vlnou pomoci parametru pred vlnou},
\eqref{C1 rychlosti za vlnou pomoci parametru pred vlnou - preznacene} in the natural conformally flat
background coordinates, namely
\begin{eqnarray}
&& x^-_i=|h'|\frac{Z_i+\bar{Z}_i}{h+\bar{h}}\,x_i^+, \nonumber\\
&& y^-_i=|h'|\frac{Z_i-\bar{Z}_i}{h-\bar{h}}\,y_i^+, \nonumber\\
&& z^-_i=|h'|\frac{|Z_i|^2-1}{|h|^2-1}\,z_i^+, \label{interakcni polohy za vlnou pomoci pred vlnou}\\
&& t^-_i\,=|h'|\frac{|Z_i|^2+1}{|h|^2+1}\,t_i^+, \nonumber
\end{eqnarray}
for positions and
\begin{eqnarray}
&& \dot{x}_i^- = a_x\dot{x}_i^+ +b_x\dot{y}_i^+ +c_x\dot{z}_i^+ +d_x\dot{t}_i^+, \nonumber \\
&& \dot{y}_i^- = a_y\dot{x}_i^+ +b_y\dot{y}_i^+ +c_y\dot{z}_i^+ +d_y\dot{t}_i^+, \nonumber \\
&& \dot{z}_i^- = a_z\dot{x}_i^+ +b_z\dot{y}_i^+ +c_z\dot{z}_i^+ +d_z\dot{t}_i^+, \label{interakcni rychlosti za vlnou pomoci pred vlnou}\\
&& \dot{t}_i^-\,=a_t\dot{x}_i^+ +b_t\dot{y}_i^+ +c_t\dot{z}_i^+ +d_t\dot{t}_i^+, \nonumber
\end{eqnarray}
for velocities. The coefficients in \eqref{interakcni polohy za vlnou pomoci pred vlnou}, \eqref{interakcni rychlosti za vlnou pomoci pred vlnou}
are somewhat complicated functions of $Z_i$, ${h\equiv h(Z_i)}$ and its derivatives ${h'\equiv h'(Z_i)}$, ${h''\equiv h''(Z_i)}$:
\begin{eqnarray}
&& a_x = \frac{1}{2}\left(c_{\eta}+c_{\bar{\eta}}+\bar{c}_{\eta}+\bar{c}_{\bar{\eta}}\right) ,\nonumber \\
&& b_x = \frac{\im }{2}\left(-c_{\eta}-c_{\bar{\eta}}+\bar{c}_{\eta}+\bar{c}_{\bar{\eta}}\right) ,\nonumber \\
&& c_x = \frac{1}{2}\left(-a_{\eta}-a_{\bar{\eta}}+b_{\eta}+b_{\bar{\eta}}\right) ,\nonumber \\
&& d_x = \frac{1}{2}\left(a_{\eta}+a_{\bar{\eta}}+b_{\eta}+b_{\bar{\eta}}\right) ,\nonumber \\
\nonumber\\
&& a_y= \frac{1}{2\,\im }\left(c_{\eta}-c_{\bar{\eta}}+\bar{c}_{\eta}-\bar{c}_{\bar{\eta}}\right) , \nonumber \\
&& b_y= \frac{1}{2}\left(-c_{\eta}+c_{\bar{\eta}}+\bar{c}_{\eta}-\bar{c}_{\bar{\eta}}\right) , \nonumber \\
&& c_y= \frac{1}{2\,\im }\left(-a_{\eta}+a_{\bar{\eta}}+b_{\eta}-b_{\bar{\eta}}\right) , \nonumber \\
&& d_y= \frac{1}{2\,\im }\left(a_{\eta}-a_{\bar{\eta}}+b_{\eta}-b_{\bar{\eta}}\right) ,
\label{explicoef}
\end{eqnarray}
\begin{eqnarray}
&& a_z = \frac{1}{2}\left(-c_{_\U}-\bar{c}_{_\U}+c_{_\V}+\bar{c}_{_\V}\right) ,\nonumber \\
&& b_z = \frac{\im }{2}\left(c_{_\U}-\bar{c}_{_\U}-c_{_\V}+\bar{c}_{_\V}\right) ,\nonumber \\
&& c_z = \frac{1}{2}\left(a_{_\U}-a_{_\V}-b_{_\U}+b_{_\V}\right) ,\nonumber \\
&& d_z = \frac{1}{2}\left(-a_{_\U}+a_{_\V}-b_{_\U}+b_{_\V}\right) ,\nonumber \\
\nonumber\\
&& a_t= \frac{1}{2}\left(c_{_\U}+\bar{c}_{_\U}+c_{_\V}+\bar{c}_{_\V}\right), \nonumber \\
&& b_t= \frac{\im }{2}\left(-c_{_\U}+\bar{c}_{_\U}-c_{_\V}+\bar{c}_{_\V}\right), \nonumber \\
&& c_t= \frac{1}{2}\left(-a_{_\U}-a_{_\V}+b_{_\U}+b_{_\V}\right), \nonumber \\
&& d_t= \frac{1}{2}\left(a_{_\U}+a_{_\V}+b_{_\U}+b_{_\V}\right), \nonumber
\end{eqnarray}
where the constants on the right-hand sides are given by expressions \eqref{acalV}, \eqref{ccalU}, \eqref{ccalbareta}.

To complete the derivation, it only remains to express the complex number $Z_i$ explicitly in terms of the initial position of the test particle
in front of the impulse. From Eqs.~\eqref{spojite interakcni parametry pomoci parametru pred vlnou} and \eqref{trmink2} it follows immediately that
${h(Z_i) = \eta_i^+\!/\,\V_i^+ = (x_i^+ +\,\im\,y_i^+)/(t_i^+ +z_i^+)}$, i.e.
\begin{equation}
Z_i = h^{-1}\!\left(\frac{x_i^+ +\,\im\,y_i^+}{t_i^+ +z_i^+}\right) , \label{Ziexplic}
\end{equation}
where $h^{-1}$ denotes the complex inverse function to $h$.

\subsection{Geometric interpretation and \\refraction formulae}
\label{subsec:refrac}

In fact, relation \eqref{Ziexplic} and its analogous counterpart in the region behind the impulse admits a nice geometric interpretation
 of the junction condition for positions across the impulse. Let us observe that from expressions \eqref{C1 polohy pred/za vlnou},
 \eqref{spojite interakcni parametry pomoci parametru pred vlnou} and \eqref{trmink2} it follows that
\begin{eqnarray}
Z_i   &=& \frac{x_i^- +\,\im\,y_i^-}{t_i^- +z_i^-} \,, \label{Z_i v kartezskych souradnicich-}\\
h(Z_i)&=& \frac{x_i^+ +\,\im\,y_i^+}{t_i^+ +z_i^+} \,. \label{Z_i v kartezskych souradnicich+}
\end{eqnarray}
Therefore, the complex mapping ${Z_i \leftrightarrow h(Z_i)}$ can be understood as an identification of the corresponding
positions of a test particle in the region behind the impulse ${(U<0)}$ and the region in front of the impulse ${(U>0)}$,
which is uniquely determined by expressions \eqref{interakcni polohy za vlnou pomoci pred vlnou}. In other words, if the
particle, moving along a geodesic, is located at ${(x_i^+,y_i^+,z_i^+)}$ when it is hit by the impulsive wave ${(U=0)}$ at the time ${t_i^+}$,
than it emerges from the impulse at the time ${t_i^-}$ at the position ${(x_i^-,y_i^-,z_i^-)}$.

\begin{figure}[ht]
\begin{center}
\includegraphics[scale=0.42]{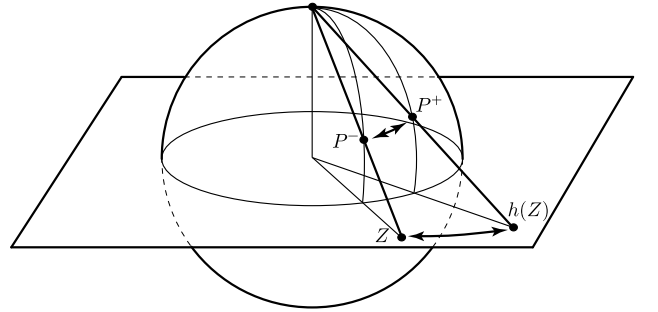}%\
%\vspace{55mm}
\end{center}
\caption{\label{figure2}%
Mapping in the complex plane $Z \leftrightarrow h(Z)$ is equivalent to identifying the points $P^-$ inside the impulsive spherical surface with the
corresponding points $P^+$ outside through the stereographic projection.
}
\end{figure}

Moreover, when the interaction time $t_i^-$ is rescaled to be equal $1$, expression~\eqref{Z_i v kartezskych souradnicich-} and its inverse
\begin{equation}
x_i^- = \frac{Z_i +\bar Z_i}{1+|Z_i|^2},\quad
y_i^- = \im\frac{\bar Z_i-Z_i}{1+|Z_i|^2},\quad
z_i^- = \frac{1-|Z_i|^2}{1+|Z_i|^2}, \label{stereoinverse}
\end{equation}
become the well-known relations for a stereographic one-to-one correspondence between a unit Riemann sphere and a complex Argand plane.
As shown in Fig.~\ref{figure2}, such mapping is obtained by projecting a straight line from the pole through $P$ onto the equatorial plane.
A point $P$ on the sphere is thus uniquely characterised by a complex number $Z$ in the complex plane (for more details see \cite{PodGri00}).

Due to the stereographic relations~\eqref{Z_i v kartezskych souradnicich-} and \eqref{Z_i v kartezskych souradnicich+}, the complex mapping
${Z_i \leftrightarrow h(Z_i)}$ thus represents a geometric identification of the points ${P^-\equiv (x_i^-,y_i^-,z_i^-)}$ and
${P^+\equiv (x_i^+,y_i^+,z_i^+)}$ on a unit sphere, which may be considered as a rescaled spherical impulsive surface ${U=0}$.
The mapping ${Z \leftrightarrow h(Z)}$ thus naturally encodes the junction conditions for position of a test particle on both sides of the impulse.

Interestingly, relations \eqref{Z_i v kartezskych souradnicich-}, \eqref{Z_i v kartezskych souradnicich+} do not involve
a cosmological constant $\Lambda$. In other words, in the conformally flat coordinates  \eqref{mink}, this geometric interpretation
is valid for expanding spherical impulses in Minkowski, de~Sitter, as well as in anti-de~Sitter space.

\begin{figure}[ht]
\begin{center}
\includegraphics[scale=0.64]{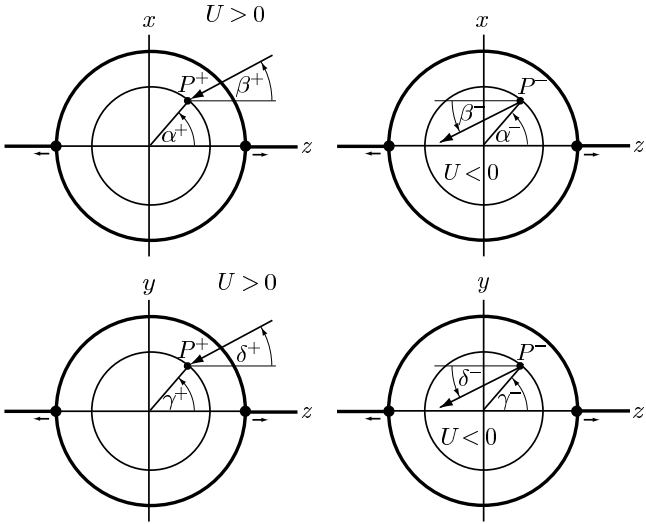}%\
\end{center}
\caption{\label{figure3}%
Definition of the angles ${\alpha, \gamma}$ characterizing position of the particle and inclination ${\beta,\delta}$ of its velocity in
the ${(x,z)}$~section (top) and ${(y,z)}$~section (bottom), respectively. Here the superscript ``$+$'' denotes quantities outside the spherical
impulse (left), while ``$-$'' labels analogous quantities inside the impulse (right). The points of interaction ${P^+ = (x_i^+,y_i^+,z_i^+)}$ and
${P^- = (x_i^-,y_i^-,z_i^-)}$ correspond to those in Fig.~\ref{figure2}. The impulsive gravitation wave is an expanding sphere indicated
in each section by the bold outer circle.}
\end{figure}

For an illustrative geometrical description of the complete effect of the spherical impulsive wave on test particles moving along geodesics
it is useful to introduce suitable \emph{angles} which characterize \emph{position} of the particle and inclination of its \emph{velocity vector}
at the instant of interaction. Specifically, in the ${(x,z)}$~section we define
\begin{equation}
\tan\alpha^{\pm} =\frac{x_i^{\pm}}{z^{\pm}_i} \,, \qquad\quad
\tan\beta^{\pm} =\frac{\dot{x}_i^{\pm}}{\dot{z}^{\pm}_i} \, ,  \label{zavedeni uhlu alpha beta}
\end{equation}
while in the perpendicular ${(y,z)}$~section we define
 \begin{equation}
\tan\gamma^{\pm}=\frac{y_i^{\pm}}{z^{\pm}_i} \,, \qquad\quad
\tan\delta^{\pm}=\frac{\dot{y}_i^{\pm}}{\dot{z}^{\pm}_i}  \,. \label{zavedeni uhlu gamma delta}
\end{equation}
The superscript ``$+$'' applies to quantities in front of the expanding impulse (outside the sphere where ${U>0}$) whereas
the superscript ``$-$'' applies to the same quantities behind the impulse (inside the sphere where ${U<0}$).
Geometrical meaning of these angles is obvious from Fig.~\ref{figure3}.

It is also useful to introduce \emph{components of the velocity} of the test particle with respect to the frames outside and inside the impulse as
\begin{equation}\label{rychlosti vzhledem k souradnemu systemu}
\big(v_x^{\pm},v_y^{\pm},v_z^{\pm}\big)\equiv\left(\frac{\dot{x}_i^{\pm}}{\dot{t}_i^{\pm}},\frac{\dot{y}_i^{\pm}}{\dot{t}_i^{\pm}},
\frac{\dot{z}_i^{\pm}}{\dot{t}_i^{\pm}}\right).
\end{equation}

If we now substitute the definitions \eqref{zavedeni uhlu alpha beta}, \eqref{zavedeni uhlu gamma delta} and
\eqref{rychlosti vzhledem k souradnemu systemu} into the equations \eqref{interakcni polohy za vlnou pomoci pred vlnou}
 and \eqref{interakcni rychlosti za vlnou pomoci pred vlnou} we obtain the following expressions which identify the positions
\begin{eqnarray}\label{obecna refrakcni formule - polohy}
&& \tan\alpha^-  \!= \frac{(|h|^2-1)\,{\mathrm{Re}}\,{Z_i}}{(|Z_i|^2-1){\mathrm{Re}}\,{h}}\tan\alpha^+ \,, \nonumber \\
&& \tan\gamma^-  \!= \frac{(|h|^2-1)\,{\mathrm{Im}}\,{Z_i}}{(|Z_i|^2-1){\mathrm{Im}}\,{h}}\tan\gamma^+ \,,
\end{eqnarray}
and inclinations of the velocity vector
\begin{eqnarray}\label{obecna refrakcni formule - rychlosti}
&& \hskip-10mm \tan\beta^-  \!= \frac{v_z^+(a_x\tan\beta^+ +b_x\tan\delta^+ +c_x) +d_x}{v_z^+(a_z\tan\beta^+ +b_z\tan\delta^+ +c_z)+d_z}\,,\nonumber\\
&& \hskip-10mm \tan\delta^- \!= \frac{v_z^+(a_y\tan\beta^+ +b_y\tan\delta^+ +c_y) +d_y}{v_z^+(a_z\tan\beta^+ +b_z\tan\delta^+ +c_z)+d_z}\,,
\end{eqnarray}
on both sides of the impulse. These explicit relations are the general \emph{refraction formulae} for motion of free test particles influenced
by the expanding impulsive gravitational wave.

\subsection{Privileged exact geodesics ${Z=\hbox{const.}}$}

In this part of Sec.~\ref{sec:geos} we restrict our attention to a privileged class of exact global geodesics given by the condition
\begin{equation}
Z=Z^0=\hbox{const.}\label{privgeod}
\end{equation}
Indeed, using the continuous form of the impulsive-wave solution \eqref{en0} it can easily be observed that
the Christoffel symbols $\,\Gamma^\mu_{UU}$, $\Gamma^\mu_{UV}$, and $\Gamma^\mu_{VV}$ vanish identically when ${\mu=Z, \bar{Z}}$.
Therefore, the geodesic equations always admit global solutions of the form \eqref{privgeod}, including across
the impulse localized at ${U=0}$ (i.e., without the necessity to \emph{assume} that the geodesics are~$C^1$).

In such a case, ${\dot{Z}_i=0}$ and expressions \eqref{C1 rychlosti pred vlnou}, \eqref{C1 rychlosti za vlnou} thus reduce to
\begin{eqnarray}
&& \dot{\V}_i^+  = A\dot{V}_i-D\dot{U}_i \,,\qquad \dot{\V}_i^-  = \frac{\dot{V}_i}{p}-\epsilon\dot{U}_i \,,\nonumber \\
&& \dot{\U}_i^+  = B\dot{V}_i-E\dot{U}_i \,,\qquad \dot{\U}_i^-  = \frac{|Z_i|^2}{p}\dot{V}_i-\dot{U}_i \,, \nonumber\\
&& \dot{\eta}_i^+= C\dot{V}_i-F\dot{U}_i \,,\qquad \ \dot{\eta}_i^-= \frac{Z_i}{p}\dot{V}_i \,,\label{C1aa}
\end{eqnarray}
respectively. Using the relations \eqref{spojite interakcni parametry pomoci parametru pred vlnou} for velocities we obtain the
(complex) constraint
\begin{equation}
\dot{\eta}_i^+\bar{C}_{,\bar{Z}}+\dot{\bar{\eta}}_i^+C_{,\bar{Z}}-\dot{\U}_i^+A_{,\bar{Z}}-\dot{\V}_i^+B_{,\bar{Z}}=0 \,, \label{podminky Z=konst.}
\end{equation}
and the following equations
\begin{eqnarray}
&& \dot{\V}_i^- = b^0_{_\V}\dot{\V}_i^+ +a^0_{_\V}\dot{\U}_i^+ +\bar{c}^0_{_\V}\dot{\eta}_i^+ +c^0_{_\V}\dot{\bar{\eta}}_i^+ , \nonumber\\
&& \dot{\U}_i^- = b^0_{_\U}\dot{\V}_i^+ +a^0_{_\U}\dot{\U}_i^+ +\bar{c}^0_{_\U}\dot{\eta}_i^+ +c^0_{_\U}\dot{\bar{\eta}}_i^+ , \label{C1 rychlosti Z0} \\
&& \dot{\eta}_i^- \,= b^0_{\eta} \dot{\V}_i^+ +a^0_{\eta}\,\dot{\U}_i^+ +\bar{c}^0_{\eta}\,\dot{\eta}_i^+ +c^0_{\eta}\,\dot{\bar{\eta}}_i^+ , \nonumber
\end{eqnarray}
where
\begin{eqnarray}
&& b^0_{_\V} = \frac{1}{p}(E-2\epsilon B)+\epsilon B \,, \nonumber \\
&& a^0_{_\V} = \frac{1}{p}(D-2\epsilon A)+\epsilon A \,, \label{komplexni koeficienty Z=konst.A} \\
&& c^0_{_\V} = -\frac{1}{p}(F-2\epsilon C)-\epsilon C \,,\nonumber
\end{eqnarray}
\begin{eqnarray}
&& b^0_{_\U} = \frac{|Z_i|^2}{p}(E-2\epsilon B)+B \,, \nonumber \\
&& a^0_{_\U} = \frac{|Z_i|^2}{p}(D-2\epsilon A)+A \,, \label{komplexni koeficienty Z=konst.B} \\
&& c^0_{_\U} = -\frac{|Z_i|^2}{p}(F-2\epsilon C)-C \,, \nonumber
\end{eqnarray}
\begin{eqnarray}
&& b^0_{\eta} = \frac{Z_i}{p}(E-2\epsilon B)\,, \nonumber \\
&& a^0_{\eta} = \frac{Z_i}{p}(D-2\epsilon A)\,,\nonumber \\
&& \bar{c}^0_{\eta} = -\frac{Z_i}{p}(\bar{F}-2\epsilon \bar{C})\,, \label{komplexni koeficienty Z=konst.C}\\
&& c^0_{\eta} = -\frac{Z_i}{p}(F-2\epsilon C)\,. \nonumber
\end{eqnarray}
The constants ${A,B,C,D,E,F,}$ and $p$ are given by the values of the functions \eqref{transe4} and \eqref{pandep} at ${Z=Z_i=Z^0}$.

In terms of the real conformally flat coordinates of metric \eqref{mink}, the velocities on both sides of the impulse are given by relations
\eqref{interakcni rychlosti za vlnou pomoci pred vlnou} where now
\begin{eqnarray}
&& a^0_x = -2\frac{\mathrm{Re}{Z_i}}{p}\,\mathrm{Re}(F-2\epsilon C),\nonumber \\
&& b^0_x = -2\frac{\mathrm{Re}{Z_i}}{p}\,\mathrm{Im}(F-2\epsilon C),\nonumber \\
&& c^0_x = \frac{\mathrm{Re}{Z_i}}{p}\,[E-D+2\epsilon(A-B)],\nonumber \\
&& d^0_x = \frac{\mathrm{Re}{Z_i}}{p}\,[E+D-2\epsilon(A+B)],\nonumber\\
&& a^0_y=  -2\frac{\mathrm{Im}{Z_i}}{p}\,\mathrm{Re}(F-2\epsilon C) , \nonumber \\
&& b^0_y=  -2\frac{\mathrm{Im}{Z_i}}{p}\,\mathrm{Im}(F-2\epsilon C) , \nonumber \\
&& c^0_y= \frac{\mathrm{Im}{Z_i}}{p}\,[E-D+2\epsilon(A-B)] , \nonumber \\
&& d^0_y= \frac{\mathrm{Im}{Z_i}}{p}\,[E+D-2\epsilon(A+B)] ,
\label{explicoef0}
\end{eqnarray}
\begin{eqnarray}
&& a^0_z = \frac{|Z_i|^2-1}{p}\,\mathrm{Re}(F-2\epsilon C)+(1-\epsilon)\,\mathrm{Re}{C},\nonumber \\
&& b^0_z = \frac{|Z_i|^2-1}{p}\,\mathrm{Im}(F-2\epsilon C)+(1-\epsilon)\,\mathrm{Im}{C} ,\nonumber \\
&& c^0_z = \frac{|Z_i|^2-1}{2p}\,[-E+D-2\epsilon(A-B)] \nonumber\\
&& \hskip20mm +{\textstyle \frac{1}{2}}(1-\epsilon)(A-B) ,\nonumber \\
&& d^0_z = \frac{|Z_i|^2-1}{2p}\,[-E-D+2\epsilon(A+B)] \nonumber\\
&& \hskip20mm -{\textstyle \frac{1}{2}}(1-\epsilon)(A+B) ,\nonumber \\
&& a^0_t= -\frac{|Z_i|^2+1}{p}\,\mathrm{Re}(F-2\epsilon C)-(1+\epsilon)\,\mathrm{Re}{C}, \nonumber \\
&& b^0_t= -\frac{|Z_i|^2+1}{p}\,\mathrm{Im}(F-2\epsilon C)-(1+\epsilon)\,\mathrm{Im}{C}, \nonumber \\
&& c^0_t= \frac{|Z_i|^2+1}{2p}\,[E-D+2\epsilon(A-B)]  \nonumber\\
&& \hskip20mm -{\textstyle \frac{1}{2}}(1+\epsilon)(A-B), \nonumber \\
&& d^0_t= \frac{|Z_i|^2+1}{2p}\,[E+D-2\epsilon(A+B)]  \nonumber\\
&& \hskip20mm +{\textstyle \frac{1}{2}}(1+\epsilon)(A+B). \nonumber
\end{eqnarray}
Moreover, from the complex constraint \eqref{podminky Z=konst.} we may express two real components of the velocity in terms of the remaining two, namely
\begin{eqnarray}
&& \dot{y}_i^+ =  \Big[\dot{x}_i^+[(B-A)\,\mathrm{Im}{F}-(E-D)\,\mathrm{Im}{C}]    \nonumber\\
&&  \qquad\qquad +2\dot{z}_i^+(\mathrm{Re}{C}\,\mathrm{Im}{F}-\mathrm{Im}{C}\,\mathrm{Re}{F})\Big] \nonumber \\
&&  \qquad\quad /[{(B-A)\,\mathrm{Re}{F}-(E-D)\,\mathrm{Re}{C}}\,],  \nonumber\\
&& \dot{t}_i^+ =  \Big[\dot{x}_i^+(BD-AE)  \\
&&  \qquad\qquad  -\dot{z}_i^+[(B+A)\,\mathrm{Re}{F}-(E+D)\,\mathrm{Re}{C}]\Big] \nonumber,\\
&&  \qquad\quad /[{(B-A)\,\mathrm{Re}{F}-(E-D)\,\mathrm{Re}{C}}\,]. \nonumber
\end{eqnarray}
Substituting these two relations and the coefficients \eqref{explicoef0} into \eqref{interakcni rychlosti za vlnou pomoci pred vlnou}, and using
Eqs.~\eqref{interakcni polohy za vlnou pomoci pred vlnou} which relate the interaction positions, we finally obtain
\begin{eqnarray}
&& \dot{x}^-_i = x_i^-\frac{(E-D)\,\dot{x}_i^++2\mathrm{Re}{F}\,\dot{z}_i^+}{(E-D)\,x_i^++2\mathrm{Re}{F}\,z_i^+} , \nonumber\\
&& \dot{y}^-_i = y_i^-\frac{(E-D)\,\dot{x}_i^++2\mathrm{Re}{F}\,\dot{z}_i^+}{(E-D)\,x_i^++2\mathrm{Re}{F}\,z_i^+}, \label{interakcni rychlosti Z=konst. final}\\
&& \dot{z}^-_i = \Big[z_i^-[(E-D)\,\dot{x}_i^++2\mathrm{Re}{F}\,\dot{z}_i^+]   \nonumber\\
&& \qquad \qquad  -(1-\epsilon)(z_i^+\dot{x}_i^+-\dot{z}_i^+x_i^+)\Big]  \nonumber\\
&& \qquad \quad /[{(E-D)\,x_i^++2\mathrm{Re}{F}\,z_i^+}], \nonumber\\
&& \dot{t}^-_i = \Big[t_i^-[(E-D)\,\dot{x}_i^++2\mathrm{Re}{F}\,\dot{z}_i^+]   \nonumber\\
&& \qquad \qquad  +(1+\epsilon)(z_i^+\dot{x}_i^+-\dot{z}_i^+x_i^+)\Big] \nonumber\\
&& \qquad \quad /[{(E-D)\,x_i^++2\mathrm{Re}{F}\,z_i^+}] . \nonumber
\end{eqnarray}
These relations are valid for any value of the cosmological constant $\Lambda$ and for an arbitrary spherical impulse. They generalize Eqs.~(4.5)
 obtained previously in \cite{PodSte03} for a special impulse generated by a snapping cosmic string in the case when ${\Lambda=0}$.

\subsection{Junction conditions in the five-dimensional representation of (anti-)de~Sitter space}

Finally, it will be illustrative to rewrite the explicit junction conditions for positions \eqref{interakcni polohy za vlnou pomoci pred vlnou} and velocities \eqref{interakcni rychlosti za vlnou pomoci pred vlnou} of test particles crossing the impulse in terms of the representation of de~Sitter or  anti-de~Sitter space as the four-dimensional hyperboloid \eqref{hyperboloid} in flat five-dimensional spacetime \eqref{5Dflatspace}. Conformally flat coordinates of the metric \eqref{mink} are obtained by the parametrization \eqref{ZcoordsCX} with \eqref{trmink2}, i.e.,
\begin{equation}
 \Z_0 = \frac{t}{\Omega}, \ \Z_1 = \frac{z}{\Omega}, \ \Z_2 = \frac{x}{\Omega}, \ \Z_3 = \frac{y}{\Omega}, \  \Z_4 = a\bigg(\frac{2}{\Omega}-1\bigg),\label{Zcoo}
\end{equation}
where ${\Omega=1+\frac1{12}\Lambda(-t^2+x^2+y^2+z^2)}$, or inversely by
\begin{equation}
 t = \frac{2a\,\Z_0}{\Z_4+a}, \ z = \frac{2a\,\Z_1}{\Z_4+a}, \ x = \frac{2a\,\Z_2}{\Z_4+a}, \ y = \frac{2a\,\Z_3}{\Z_4+a},\label{Zcooinv}
\end{equation}
with ${\Omega=2a/(\Z_4+a)}$.

As explained in Sec.~\ref{sec:eiw}, the expanding spherical impulse located at ${U=0}$ corresponds to the cut ${\Z_4=a}$ through the hyperboloid, see Fig.~\ref{figure1}. Therefore, at the instant of interaction the particle is located at
\begin{equation}
 t_i = \Z_{0i}, \  z_i = \Z_{1i}, \  x_i = \Z_{2i}, \ y_i = \Z_{3i}, \quad  \Omega_i=1.\label{Zcoointer}
\end{equation}
The junction conditions \eqref{interakcni polohy za vlnou pomoci pred vlnou} for positions thus imply
\begin{eqnarray}
&& \Z_{0i}^-=|h'|\frac{|Z_i|^2+1}{|h|^2+1}\,\Z_{0i}^+\,, \nonumber\\
&& \Z_{1i}^-=|h'|\frac{|Z_i|^2-1}{|h|^2-1}\,\Z_{1i}^+\,, \nonumber\\
&& \Z_{2i}^-=|h'|\frac{Z_i+\bar{Z}_i}{h+\bar{h}}\,\Z_{2i}^+\,, \label{interakcni polohyZ}\\
&& \Z_{3i}^-=|h'|\frac{Z_i-\bar{Z}_i}{h-\bar{h}}\,\Z_{3i}^+\,, \nonumber\\
&& \Z_{4i}^- \,=\, a\,=\ \Z_{4i}^+\,.  \nonumber
\end{eqnarray}

By differentiating Eqs.~\eqref{Zcoo} and evaluating them at the interaction time we obtain the relations
\begin{eqnarray}
&& \dot{\Z}_{0i}=\dot{t}_i -\dot{\Omega}_i t_i \,, \nonumber \\
&& \dot{\Z}_{1i}=\dot{z}_i -\dot{\Omega}_i z_i \,, \nonumber \\
&& \dot{\Z}_{2i}=\dot{x}_i -\dot{\Omega}_i x_i \,, \label{interakcni rychlostiZ}\\
&& \dot{\Z}_{3i}=\dot{y}_i -\dot{\Omega}_i y_i \,, \nonumber \\
&& \dot{\Z}_{4i}=-2a\dot{\Omega}_i \,, \nonumber
\end{eqnarray}
where ${\dot{\Omega}_i=\frac1{6}\Lambda(-t_i\dot{t}_i+x_i\dot{x}_i+y_i\dot{y}_i+z_i\dot{z}_i)}$, which are valid both in front and behind the impulse.
From expressions \eqref{interakcni rychlosti za vlnou pomoci pred vlnou} and \eqref{Zcoointer}
we thus obtain the following relations between velocities on both sides of the impulse,
\begin{eqnarray}
&& \dot{\Z}_{\p i}^{-}  = a_\p\dot{\Z}_{2i}^{+}+b_\p\dot{\Z}_{3i}^{+}+c_\p\dot{\Z}_{1i}^{+}+d_\p\dot{\Z}_{0i}^{+}+n_\p\dot{\Z}_{4i}^{+}, \nonumber \\
&& \dot{\Z}_{4i}^{-} =-2a\dot{\Omega}_i=\dot{\Z}_{4i}^{+}, \label{navazani rychlostiZ}
\end{eqnarray}
were we denoted ${\p=0,1,2,3}$. The constant coefficients ${(a_0,a_1,a_2,a_3)}\equiv {(a_t,a_z,a_x,a_y)}$, and similarly ${b_\p,c_\p,d_\p}$,
are given by \eqref{explicoef}. The coefficients $n_\p$  are defined as
\begin{equation}
n_\p=-\frac{1}{2a}\left(a_\p \Z_{2i}^{+}+b_\p \Z_{3i}^{+}+c_\p \Z_{1i}^{+}+d_\p \Z_{0i}^{+}-\Z_{\p i}^{-}\right),
\label{omega}
\end{equation}
where $\Z_{\p i}^{-}$ should be expressed using \eqref{interakcni polohyZ}. Relations \eqref{navazani rychlostiZ} can also be written in the matrix form:
\begin{equation}
\left(\begin{array}{ccccc@{\ }r}
\dot{\Z}_{2i}^{-} \\
\dot{\Z}_{3i}^{-} \\
\dot{\Z}_{1i}^{-} \\
\dot{\Z}_{0i}^{-} \\
\dot{\Z}_{4i}^{-}
\end{array}\right)
=
\left(\begin{array}{ccccc@{\ }r}
a_x & b_x & c_x & d_x & n_x \\
a_y & b_y & c_y & d_y & n_y \\
a_z & b_z & c_z & d_z & n_z \\
a_t & b_t & c_t & d_t & n_t \\
0  & 0  & 0  & 0  & 1
\end{array}\right)
\left(\begin{array}{ccccc@{\ }r}
\dot{\Z}_{2i}^{+} \\
\dot{\Z}_{3i}^{+} \\
\dot{\Z}_{1i}^{+} \\
\dot{\Z}_{0i}^{+} \\
\dot{\Z}_{4i}^{+}
\end{array}\right).
\label{navazani_rychlosti_5D}
\end{equation}
Expressions \eqref{interakcni polohyZ} and \eqref{navazani_rychlosti_5D} are explicit junction conditions which relate the positions and velocities
of test particles when they cross an expanding spherical impulse. They are expressed in the natural five-dimensional coordinates of
constant-curvature spaces with ${\Lambda\not=0}$, namely the (anti-)de~Sitter half-space in front of the impulse, and the analogous
half-space behind it. Obviously, the junction conditions depend on the complex function $h(Z)$ which defines the specific impulse of this type.

The advantage of expressing the junction conditions for geodesics in the ``geometrical'' five-dimensional formalism is that they may easily be applied to obtain the corresponding explicit
conditions in terms of \emph{any} standard coordinates of de Sitter or anti-de~Sitter background space. We will demonstrate this procedure in the next section in which we concentrate of spherical impulses generated by  a snapping cosmic string. Their influence on particles will most naturally be expresses in global coordinates in de~Sitter space with a synchronous time coordinate, see Sec.~\ref{sec:string}.~C.

\section{Geodesics crossing the impulse generated by a snapped cosmic string}
\label{sec:string}

\begin{figure}[ht]
\begin{center}
\includegraphics[scale=0.48]{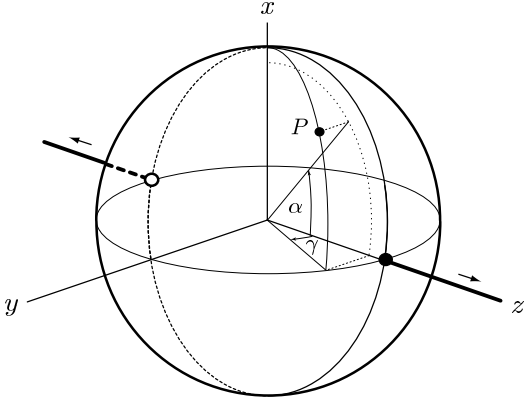}%\
\vspace{-2mm}
\end{center}
\caption{\label{figure4}%
Geometry of a spherical impulse expanding with the speed of light. It is generated by a snapped cosmic string, whose remnants are
two semi-infinite strings located along the $z$~axis outside the impulsive wave. Any point $P$ on the impulse is described
by two angles $\alpha$ and $\gamma$ which characterize its projections to the $(x,z)$ and $(y,z)$ planes, respectively (cf.~Fig.~\ref{figure3}).
}
\end{figure}

The general results obtained above will now be applied to an important particular family of spacetimes in which the expanding spherical impulsive wave is
generated by a snapped cosmic string (identified by a deficit angle in the region ${U>0}$ in front of the impulse). Such exact vacuum solutions were introduced and discussed in a number of works, e.g.~\cite{NutPen92,BicSch89,Bicak90,GlePul89,PodGri00,PodGri01a,PodGri01b,PodGri04}. These can be written in the form of the metric \eqref{en0} with
\begin{equation}
H(Z)=\frac{\frac{1}{2}\delta(1-\frac{1}{2}\delta)}{Z^2}, \label{string}
\end{equation}
which is obtained from the complex function
\begin{equation}
h(Z)=Z^{1-\delta}, \label{hstring}
\end{equation}
using the expression \eqref{Schwarz}. Here ${\delta\in [0,1)}$ is a real constant which characterizes the deficit angle $2\pi\delta$
of the snapped string that is located in the region outside the impulse along the $z$~axis given by ${\eta=0}$, as shown in Fig.~\ref{figure4}
(see \cite{NutPen92,PodGri00} for more details).

\subsection{Explicit junction conditions}
\label{subsec:explstringcond}

Expressions~\eqref{interakcni polohy za vlnou pomoci pred vlnou} which are the junction conditions for positions (in the natural conformally flat background coordinates on both sides of the impulse) are thus
\begin{eqnarray}
&& x^-_i=(1-\delta)\, |Z_i|^{-\delta}\frac{Z_i+\bar{Z}_i}{Z_i^{1-\delta}+\bar Z_i^{1-\delta}}\,x_i^+, \nonumber\\
&& y^-_i=(1-\delta)\, |Z_i|^{-\delta}\frac{Z_i-\bar{Z}_i}{Z_i^{1-\delta}-\bar Z_i^{1-\delta}}\,y_i^+, \nonumber\\
&& z^-_i=(1-\delta)\, \frac{|Z_i|-|Z_i|^{-1}}{|Z_i|^{1-\delta}-|Z_i|^{\delta-1}}\,z_i^+, \label{interakcni polohy pro strunu}\\
&& t^-_i\,=(1-\delta)\, \frac{|Z_i|+|Z_i|^{-1}}{|Z_i|^{1-\delta}+|Z_i|^{\delta-1}}\,t_i^+, \nonumber
\end{eqnarray}
where, in view of relation~\eqref{Ziexplic},
\begin{equation}
Z_i^{1-\delta} =\frac{x_i^+ +\,\im\,y_i^+}{t_i^+ +z_i^+}\,. \label{Ziexplicstruna}
\end{equation}
Let us also recall that ${(x_i^\pm)^2 + (y_i^\pm)^2 + (z_i^\pm)^2 = (t_i^\pm)^2}$ because the positions are evaluated on the impulse ${U=0}$.

Similarly, it is straightforward to evaluate the specific form of the coefficients \eqref{explicoef} which relate the velocities in Eqs.~\eqref{interakcni rychlosti za vlnou pomoci pred vlnou}, namely
\begin{eqnarray}
&& a_x= \frac{|Z_i|^\delta}{2(1-\delta)}\Big[(1-\dpul)^2(Z_i^{-\delta}+\bar Z_i^{-\delta})\nonumber \\
&&     \hspace*{22mm}+\ddctvrt|Z_i|^{-2}(Z_i^{2-\delta}+\bar Z_i^{2-\delta})\Big], \nonumber \\
&& b_x= \frac{|Z_i|^\delta}{2\im(1-\delta)}\Big[(1-\dpul)^2(Z_i^{-\delta}-\bar Z_i^{-\delta})\nonumber \\
&&     \hspace*{22mm}+\ddctvrt|Z_i|^{-2}(Z_i^{2-\delta}-\bar Z_i^{2-\delta})\Big], \label{abcdStringxgen} \\
&& c_x = \frac{\delta(1-\dpul)}{4(1-\delta)}|Z_i|^\delta \Big[(Z_i^{-1}+\bar Z_i^{-1})\nonumber \\
&&     \hspace*{32mm}-|Z_i|^{-2\delta} (Z_i+\bar Z_i)\Big], \nonumber \\
&& d_x = -\frac{\delta(1-\dpul)}{4(1-\delta)}|Z_i|^\delta \Big[(Z_i^{-1}+\bar Z_i^{-1})\nonumber \\
&&     \hspace*{32mm}+|Z_i|^{-2\delta} (Z_i+\bar Z_i)\Big], \nonumber
\end{eqnarray}
\begin{eqnarray}
&& a_y= \frac{\im\,|Z_i|^\delta}{2(1-\delta)}\Big[(1-\dpul)^2(Z_i^{-\delta}-\bar Z_i^{-\delta})\nonumber \\
&&     \hspace*{22mm}-\ddctvrt|Z_i|^{-2}(Z_i^{2-\delta}-\bar Z_i^{2-\delta})\Big], \nonumber \\
&& b_y= \frac{|Z_i|^\delta}{2(1-\delta)}\Big[(1-\dpul)^2(Z_i^{-\delta}+\bar Z_i^{-\delta})\nonumber \\
&&     \hspace*{22mm}-\ddctvrt|Z_i|^{-2}(Z_i^{2-\delta}+\bar Z_i^{2-\delta})\Big], \label{abcdStringygen} \\
&& c_y = \frac{\im\delta(1-\dpul)}{4(1-\delta)}|Z_i|^\delta \Big[(Z_i^{-1}-\bar Z_i^{-1})\nonumber \\
&&     \hspace*{32mm}+|Z_i|^{-2\delta} (Z_i-\bar Z_i)\Big], \nonumber \\
&& d_y = \frac{\delta(1-\dpul)}{4\im(1-\delta)}|Z_i|^\delta \Big[(Z_i^{-1}-\bar Z_i^{-1})\nonumber \\
&&     \hspace*{32mm}-|Z_i|^{-2\delta} (Z_i-\bar Z_i)\Big], \nonumber
\end{eqnarray}
\begin{eqnarray}
&& a_z= \frac{\delta(1-\dpul)}{4(1-\delta)}|Z_i|^\delta \Big[(Z_i^{1-\delta}+\bar Z_i^{1-\delta})\nonumber \\
&&     \hspace*{27mm}-|Z_i|^{-2}(Z_i^{1-\delta}+\bar Z_i^{1-\delta})\Big], \nonumber \\
&& b_z= \frac{\delta(1-\dpul)}{4\im (1-\delta)}|Z_i|^\delta \Big[(Z_i^{1-\delta}-\bar Z_i^{1-\delta})\nonumber \\
&&     \hspace*{27mm}-|Z_i|^{-2}(Z_i^{1-\delta}-\bar Z_i^{1-\delta})\Big], \label{abcdStringzgen} \\
&& c_z= \frac{1}{2(1-\delta)}\Big[(1-\dpul)^2(|Z_i|^\delta+|Z_i|^{-\delta}) \nonumber \\
&&     \hspace*{22mm}-\ddctvrt(|Z_i|^{2-\delta}+|Z_i|^{\delta-2})\Big], \nonumber \\
&& d_z= \frac{-1}{2(1-\delta)}\Big[(1-\dpul)^2(|Z_i|^\delta-|Z_i|^{-\delta}) \nonumber \\
&&     \hspace*{22mm}+\ddctvrt(|Z_i|^{2-\delta}-|Z_i|^{\delta-2})\Big], \nonumber
\end{eqnarray}
\begin{eqnarray}
&& a_t= -\frac{\delta(1-\dpul)}{4(1-\delta)}|Z_i|^\delta \Big[(Z_i^{1-\delta}+\bar Z_i^{1-\delta})\nonumber \\
&&     \hspace*{27mm}+|Z_i|^{-2}(Z_i^{1-\delta}+\bar Z_i^{1-\delta})\Big], \nonumber \\
&& b_t= \frac{\im \delta(1-\dpul)}{4(1-\delta)}|Z_i|^\delta \Big[(Z_i^{1-\delta}-\bar Z_i^{1-\delta})\nonumber \\
&&     \hspace*{27mm}+|Z_i|^{-2}(Z_i^{1-\delta}-\bar Z_i^{1-\delta})\Big],  \label{abcdStringtgen} \\
&& c_t= \frac{-1}{2(1-\delta)}\Big[(1-\dpul)^2(|Z_i|^\delta-|Z_i|^{-\delta}) \nonumber\\
&&     \hspace*{22mm}-\ddctvrt (|Z_i|^{2-\delta}- |Z_i|^{\delta-2})\Big], \nonumber \\
&& d_t=  \frac{1}{2(1-\delta)}\Big[(1-\dpul)^2(|Z_i|^\delta+|Z_i|^{-\delta}) \nonumber\\
&&     \hspace*{22mm}+\ddctvrt (|Z_i|^{2-\delta}+ |Z_i|^{\delta-2})\Big]. \nonumber
\end{eqnarray}

Considering the structure of these relations, it is very convenient to reparametrize the complex number $Z_i$ in the polar form as
\begin{equation}
Z_i \equiv R\, e^{\im\, \Phi}, \label{plarparZi}
\end{equation}
where ${R=|Z_i|}$ and $\Phi$ are constants representing its modulus and phase, respectively. It immediately follows from the relation~\eqref{Ziexplicstruna} that
\begin{eqnarray}
&& R=\left(\frac{(x_i^+)^2 + (y_i^+)^2}{(t_i^+ + z_i^+)^2}\right)^\frac{1}{2(1-\delta)}=\left(\frac{t_i^+ - z_i^+}{t_i^+ + z_i^+}\right)^\frac{1}{2(1-\delta)}, \nonumber\\
&& \tan\big((1-\delta)\Phi\big)=\frac{y_i^+}{x_i^+}\,. \label{RPhipomocixyzt}
\end{eqnarray}
The junction conditions~\eqref{interakcni polohy pro strunu} for positions then take the form
\begin{eqnarray}
&& x^-_i=(1-\delta)\, \frac{\cos\Phi}{\cos\big((1-\delta)\Phi\big)}\,x_i^+, \nonumber\\
&& y^-_i=(1-\delta)\, \frac{\sin\Phi}{\sin\big((1-\delta)\Phi\big)}\,y_i^+, \nonumber\\
&& z^-_i=(1-\delta)\, \frac{R-R^{-1}}{R^{1-\delta}-R^{\delta-1}}\,z_i^+, \nonumber\\
&& t^-_i\,=(1-\delta)\, \frac{R+R^{-1}}{R^{1-\delta}+R^{\delta-1}}\,t_i^+, \label{interakcni polohy pro strunup}
\end{eqnarray}
and the coefficients \eqref{abcdStringxgen}--\eqref{abcdStringtgen} simplify to
\begin{eqnarray}
&& a_x=  \frac{1}{1-\delta}\Big[(1-\dpul)^2\cos(\delta\Phi)+\ddctvrt\cos\big((2-\delta)\Phi\big)\Big],\nonumber \\
&& b_x= \frac{-1}{1-\delta}\Big[(1-\dpul)^2\sin(\delta\Phi)-\ddctvrt\sin\big((2-\delta)\Phi\big)\Big],\nonumber \\
&& c_x =  \frac{\delta(1-\dpul)}{2(1-\delta)}\big(R^{\delta-1}-R^{1-\delta}\big)\cos\Phi,\nonumber\\
&& d_x = -\frac{\delta(1-\dpul)}{2(1-\delta)}\big(R^{\delta-1}+R^{1-\delta}\big)\cos\Phi, \label{abcdStringxgenp}
\end{eqnarray}
\begin{eqnarray}
&& a_y= \frac{1}{1-\delta}\Big[(1-\dpul)^2\sin(\delta\Phi)+\ddctvrt\sin\big((2-\delta)\Phi\big)\Big],\nonumber \\
&& b_y= \frac{1}{1-\delta}\Big[(1-\dpul)^2\cos(\delta\Phi)-\ddctvrt\cos\big((2-\delta)\Phi\big)\Big],\nonumber \\
&& c_y =  \frac{\delta(1-\dpul)}{2(1-\delta)}\big(R^{\delta-1}-R^{1-\delta}\big)\sin\Phi,\nonumber\\
&& d_y = -\frac{\delta(1-\dpul)}{2(1-\delta)}\big(R^{\delta-1}+R^{1-\delta}\big)\sin\Phi, \label{abcdStringygenp}
\end{eqnarray}
\begin{eqnarray}
&& a_z= \frac{\delta(1-\dpul)}{2(1-\delta)}\big(R-R^{-1}\big)\cos\big((1-\delta)\Phi\big),\nonumber \\
&& b_z= \frac{\delta(1-\dpul)}{2(1-\delta)}\big(R-R^{-1}\big)\sin\big((1-\delta)\Phi\big),\nonumber \\
&& c_z = \frac{1}{2(1-\delta)}\Big[(1-\dpul)^2(R^\delta+R^{-\delta}) \nonumber \\
&&     \hspace*{22mm} -\ddctvrt(R^{2-\delta}+R^{\delta-2})\Big],\nonumber\\
&& d_z = \frac{-1}{2(1-\delta)}\Big[(1-\dpul)^2(R^\delta-R^{-\delta}) \nonumber \\
&&     \hspace*{22mm} +\ddctvrt(R^{2-\delta}-R^{\delta-2})\Big], \label{abcdStringzgenp}
\end{eqnarray}
\begin{eqnarray}
&& a_t= -\frac{\delta(1-\dpul)}{2(1-\delta)}\big(R+R^{-1}\big)\cos\big((1-\delta)\Phi\big),\nonumber \\
&& b_t= -\frac{\delta(1-\dpul)}{2(1-\delta)}\big(R+R^{-1}\big)\sin\big((1-\delta)\Phi\big),\nonumber \\
&& c_t = \frac{-1}{2(1-\delta)}\Big[(1-\dpul)^2(R^\delta-R^{-\delta}) \nonumber \\
&&     \hspace*{22mm} -\ddctvrt(R^{2-\delta}-R^{\delta-2})\Big],\nonumber\\
&& d_t = \frac{1}{2(1-\delta)}\Big[(1-\dpul)^2(R^\delta+R^{-\delta}) \nonumber \\
&&     \hspace*{22mm} +\ddctvrt(R^{2-\delta}+R^{\delta-2})\Big]. \label{abcdStringtgenp}
\end{eqnarray}
Notice finally that the terms involving $R$ could also be conveniently expressed using the hyperbolic functions as
\begin{eqnarray}
R-R^{-1} &=& 2\sinh r,\nonumber \\
R+R^{-1} &=& 2\cosh r,\nonumber \\
R^{\delta}-R^{-\delta} &=& 2\sinh (\delta r),\nonumber \\
R^{\delta}+R^{-\delta} &=& 2\cosh (\delta r),\label{hyperb}\\
R^{1-\delta}-R^{\delta-1} &=& 2\sinh \big((1-\delta)r\big),\nonumber \\
R^{1-\delta}+R^{\delta-1} &=& 2\cosh \big((1-\delta)r\big),\nonumber \\
R^{2-\delta}-R^{\delta-2} &=& 2\sinh \big((2-\delta)r\big),\nonumber \\
R^{2-\delta}+R^{\delta-2} &=& 2\cosh \big((2-\delta)r\big),\nonumber
\end{eqnarray}
where
\begin{equation}
r\equiv\log R  =\frac{1}{2(1-\delta)}\log \left(\frac{t_i^+ - z_i^+}{t_i^+ + z_i^+}\right). \label{rlogR}
\end{equation}
Employing the relation ${ t_i^+=\sqrt{(x_i^+)^2 +(y_i^+)^2 + (z_i^+)^2 }}={z_i^+\sqrt{1+\tan^2\alpha^+ +\tan^2\gamma^+}}$, this can be written
explicitly in terms of the initial position as
\begin{equation}
r=\frac{1}{2(1-\delta)}\log \left(\frac{\sqrt{1+\tan^2\alpha^+ +\tan^2\gamma^+} - 1}{\sqrt{1+\tan^2\alpha^+ +\tan^2\gamma^+} + 1}\right). \label{rlogRplus}
\end{equation}
Moreover,
\begin{equation}
\tan\big((1-\delta)\Phi\big)=\frac{\tan\gamma^+}{\tan\alpha^+}. \label{RPhiplus}
\end{equation}
The above formulae enable us to investigate behaviour of arbitrary geodesics which cross the spherical impulse generated by a snapped cosmic string.

\subsection{Analysis and description of the resulting motion}
\label{subsec:analystring}

For simplicity, let us consider a family of test particles which are \emph{at rest} in front of the impulse (i.e., in the  constant-curvature region ${U>0}$). Specifically, we will first assume that the velocities of the particles in the coordinates~\eqref{mink} of Minkowski, de~Sitter or anti-de~Sitter space vanish, ${\dot{x}^+=\dot{y}^+=\dot{z}^+=0}$.

Junction conditions~\eqref{interakcni rychlosti za vlnou pomoci pred vlnou} for the velocities across the impulse thus simplify considerably to
\begin{eqnarray}
&& \dot{x}_i^- = d_x\,\dot{t}_i^+, \nonumber \\
&& \dot{y}_i^- = d_y\,\dot{t}_i^+, \nonumber \\
&& \dot{z}_i^- = d_z\,\dot{t}_i^+, \label{interakcni rychlosti string}\\
&& \dot{t}_i^-\,=d_t\,\dot{t}_i^+, \nonumber
\end{eqnarray}
where the constants ${d_x, d_y, d_z, d_t}$ are given by~\eqref{abcdStringxgenp}, \eqref{abcdStringygenp}, \eqref{abcdStringzgenp}, \eqref{abcdStringtgenp}, respectively. Using the definitions~\eqref{zavedeni uhlu alpha beta}, \eqref{zavedeni uhlu gamma delta} and relations~\eqref{interakcni polohy pro strunup} for positions, it is straightforward to obtain the following  refraction formulae:
\begin{eqnarray}
\hskip-4mm \tan\alpha^- &=& \frac{\sinh \big((1-\delta)r\big)}{\sinh r} \frac{\cos\Phi}{\cos\big((1-\delta)\Phi\big)}\,\tan\alpha^+,  \label{stringrefx}\\
\hskip-4mm \tan\beta^- &=&\frac{\delta(1-\dpul)\cosh \big((1-\delta)r\big)\cos\Phi}{(1-\dpul)^2\sinh (\delta r)+\ddctvrt\sinh \big((2-\delta)r\big)},\nonumber
\end{eqnarray}
and
\begin{eqnarray}
\hskip-4mm \tan\gamma^- &=& \frac{\sinh \big((1-\delta)r\big)}{\sinh r} \frac{\sin\Phi}{\sin\big((1-\delta)\Phi\big)}\,\tan\gamma^+,  \label{stringrefy}\\
\hskip-4mm \tan\delta^- &=&\frac{\delta(1-\dpul)\cosh \big((1-\delta)r\big)\sin\Phi}{(1-\dpul)^2\sinh (\delta r)+\ddctvrt\sinh \big((2-\delta)r\big)}.\nonumber
\end{eqnarray}
\begin{figure}[t]
\begin{center}
\includegraphics[scale=0.78]{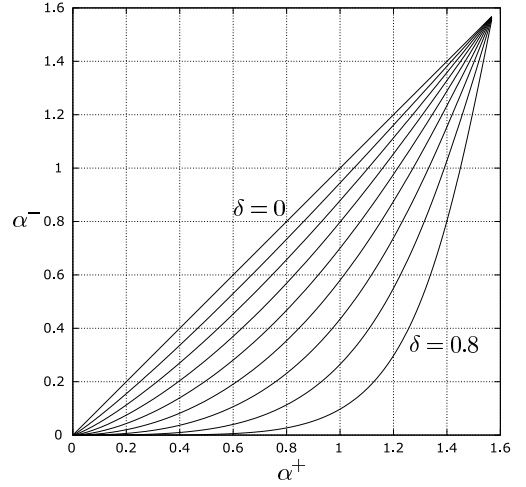}%\
\vspace{-7mm}
\end{center}
\caption{\label{figure5}%
The function ${\alpha^-(\alpha^+)}$ which determines the displacement of the position of a particle when it crosses the impulse generated by a snapped cosmic string.
The curves correspond to different values of the deficit angle parameter ${\delta=0, 0.1,0.2,\ldots,0.8}$.
}
\end{figure}
\begin{figure}[h]
\begin{center}
\includegraphics[scale=0.78]{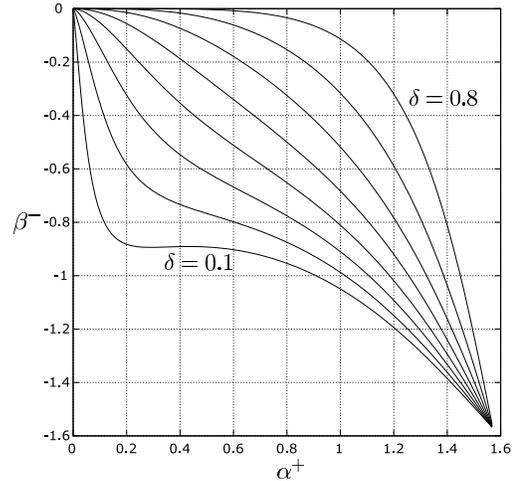}%\
\vspace{-7mm}
\end{center}
\caption{\label{figure6}%
The function ${\beta^-(\alpha^+)}$ which determines the dependence of the velocity vector inclination behind the impulse on the particle's
position in front of the impulse. The curves plotted correspond to ${\delta=0.1,0.2,\ldots,0.8}$.
}
\end{figure}
Due to the axial symmetry of the spacetime along the $z$~axis (where the string is located in front of the impulse) it is natural to restrict
attention to a \emph{ring of test particles located in the} ${(x^+,z^+)}$~plane, i.e., assuming ${y^+=0}$. From~\eqref{zavedeni uhlu gamma delta} it
follows that ${\gamma^+=0}$ and, using~\eqref{RPhiplus}, this implies ${\Phi=0}$. Consequently, Eqs.~\eqref{stringrefy} reduce to
${\gamma^-=0=\delta^-}$. It follows that ${y_i^-=0=\dot{y}_i^-}$, and motion of such particles will thus remain in the $(x^-,z^-)$~plane behind the impulse.
Relations~\eqref{stringrefx}, which describe the motion in the ${(x,z)}$~plane, now reduce to
\begin{eqnarray}
\hskip-4mm \tan\alpha^- &=& \frac{\sinh \big((1-\delta)r\big)}{\sinh r} \,\tan\alpha^+,  \label{stringrefxy=0}\\
\hskip-4mm \tan\beta^- &=&\frac{\delta(1-\dpul)\cosh \big((1-\delta)r\big)}{(1-\dpul)^2\sinh (\delta r)+\ddctvrt\sinh \big((2-\delta)r\big)},\nonumber
\end{eqnarray}
where the parameter $r$ is given by Eq.~\eqref{rlogRplus}. Because ${\gamma^+=0}$, this further simplifies to
\begin{equation}
r=\frac{1}{1-\delta}\log \left(\tan\frac{\alpha^+}{2}\right). \label{rlogRy=o}
\end{equation}

It is now possible to visualize the effect of the impulse generated by a snapped cosmic string on such a ring of test particles by
plotting the corresponding graphs. In Fig.~\ref{figure5} and Fig.~\ref{figure6} we draw the functions ${\alpha^-(\alpha^+)}$ and ${\beta^-(\alpha^+)}$,
respectively, which are given by~\eqref{stringrefxy=0} with \eqref{rlogRy=o}, for several discrete values of the parameter $\delta$. The geometrical meaning
of these angles is described in Figs.~\ref{figure3} and~\ref{figure4}. The angle ${\alpha^+}$ parameterizes position of a particle of the ring in front of the impulse,
while  ${\alpha^-}$ and ${\beta^-}$ determine, respectively, its position and velocity vector inclination behind the impulse.

\begin{figure}[t]
\begin{center}
\includegraphics[scale=0.93]{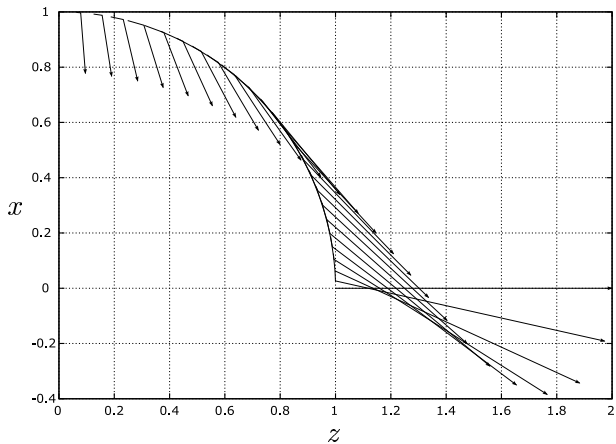}%\
\vspace{-5mm}
\end{center}
\caption{\label{figure7}%
The effect of the impulse with ${\delta=0.2}$ on a ring of initially static test particles in the $(x,z)$~plane. The particles are shifted and they start to move, as
indicated by their velocity vectors with components ${(v_x^-,v_z^-)}$ behind the impulse. The impulse is scaled here in such a way that it is given by a  unit sphere on both sides of the impulse.}
\end{figure}

\begin{figure}[t]
\begin{center}
\includegraphics[scale=0.93]{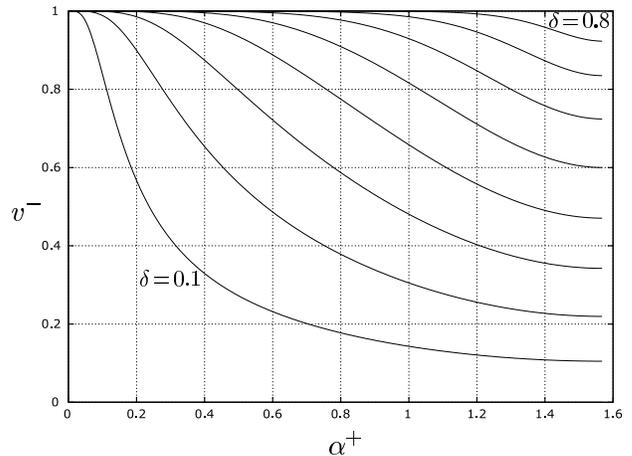}%\
\vspace{-5mm}
\end{center}
\caption{\label{figure8}%
The magnitude $v^-$ of the velocity vector behind the impulse as a function of particle's initial position $\alpha^+$.
The curves plotted correspond to different values of the parameter ${\delta=0.1, 0.2,\ldots, 0.8}$.}
\end{figure}

\begin{figure}[h]
\begin{center}
\includegraphics[scale=0.93]{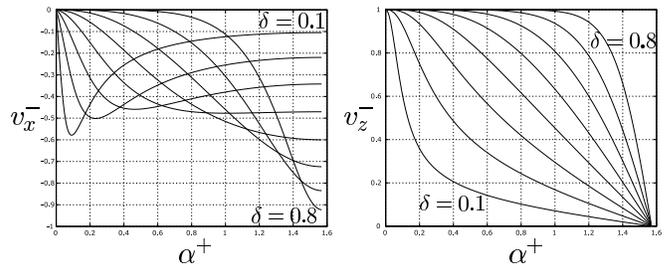}%\
\vspace{-5mm}
\end{center}
\caption{\label{figure9}%
The components ${v_x^-}$ (left) and ${v_z^-}$ (right) of the velocity vector behind the impulse as a function of initial position $\alpha^+$.
The curves correspond to ${\delta=0.1, 0.2,\ldots, 0.8}$.}
\end{figure}

Combining these two relations, we plot in Fig.~\ref{figure7} the motion of the (initially static) ring caused by the impulse.
It can be seen that the particles are displaced towards the string, and directions of their velocities are oriented ``along'' the string.
Particles located close to the string in front the impulse are accelerated almost to the speed of light behind the impulse, and are
``dragged'' along the string (except those in the perpendicular plane ${z=0}$ corresponding to ${\alpha^+=\frac{\pi}{2}}$). In Fig.~\ref{figure8} we plot the magnitude ${v^-=\sqrt{(v_x^-)^2+(v_z^-)^2}}$ of the resulting velocity vector
 as a function of ${\alpha^+}$ for several values of the parameter $\delta$. Indeed, for small
values of the angle ${\alpha^+}$ the speed approaches that of light, ${v^-\to1}$. The components ${v_x^-=(\dot{x}_i^-/\dot{t}_i^-)=d_x/d_t}$ and ${v_z^-=(\dot{z}_i^-/\dot{t}_i^-)=d_z/d_t}$
are separately drawn in Fig.~\ref{figure9}. Since ${v_x^-(0)=0}$, ${v_z^-(0)=1}$ for any ${\delta>0}$, the
particles close to the string are accelerated ``parallelly'' along it. For ${\delta\to0}$, ${v_z^-(\alpha^+)}$ becomes zero
everywhere except at ${\alpha^+=0}$ where the string is located. Also, ${v_z^-(\frac{\pi}{2})=0}$ which means that
the particles in the transverse plane ${z=0}$ are accelerated ``perpendicularly'' and they thus stay in this plane,
which is consistent with the symmetry of the system. The velocity vectors corresponding to such components are indicated in Fig.~\ref{figure7} by arrows.


\begin{figure}[t]
\begin{center}
\includegraphics[scale=0.93]{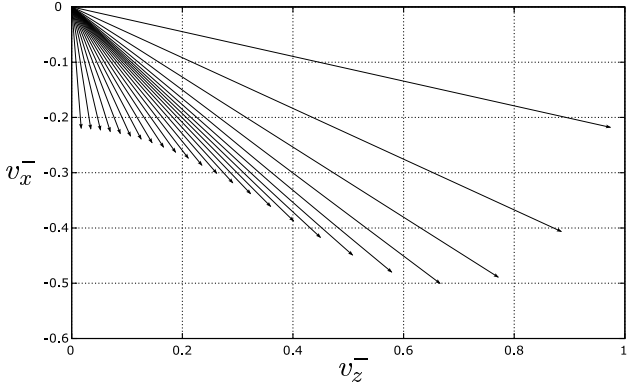}%\
\vspace{-5mm}
\end{center}
\caption{\label{figure10}%
The velocity vectors for ${\delta=0.2}$ plotted as a function of the initial position $\alpha^+$ of the particle in the ring.}
\end{figure}

\begin{figure}[h]
\begin{center}
\includegraphics[scale=0.93]{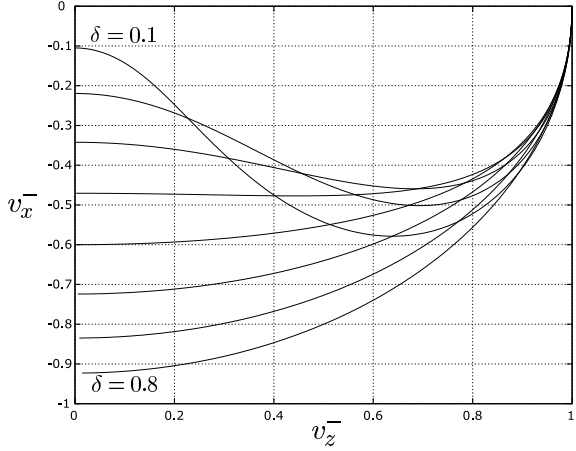}%\
\vspace{-5mm}
\end{center}
\caption{\label{figure11}%
Envelope of the velocity vectors for all ${\alpha^+\in[0,\frac{\pi}{2}]}$, plotted for ${\delta=0.1, 0.2,\ldots, 0.8}$.}
\end{figure}

To compare these velocity vectors for different values of the initial position $\alpha^+$, we plot them in Fig.~\ref{figure10} from the common origin.
The endpoints of these arrows for all ${\alpha^+\in[0,\frac{\pi}{2}]}$ form a smooth curve, which is drawn in Fig.~\ref{figure11} for several discrete 
values of the conicity parameter~$\delta$. For small $\delta$ there is a single minimum in such curves, while for large values of $\delta$ the curves approach a unit circle since the particles are accelerated by the impulse almost to the speed of light in all directions.

\begin{figure}[h]
\begin{center}
\includegraphics[scale=0.90]{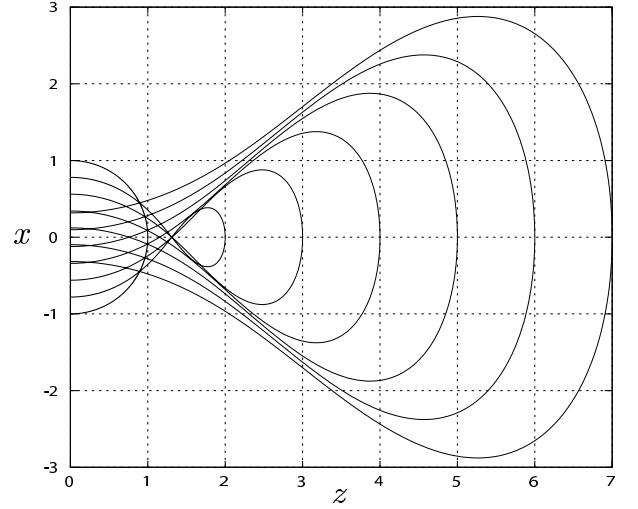}%\
\vspace{-5mm}
\end{center}
\caption{\label{figure12}%
Time sequence showing the deformation of the ring of test particles (indicated here by an initial semi-circle of unit radius for ${\alpha^+\in[-\frac{\pi}{2},\frac{\pi}{2}]}$) caused by the spherical impulse generated by a snapping cosmic string  with ${\delta=0.2}$.}
\end{figure}

\begin{figure}[h]
\begin{center}
\includegraphics[scale=1]{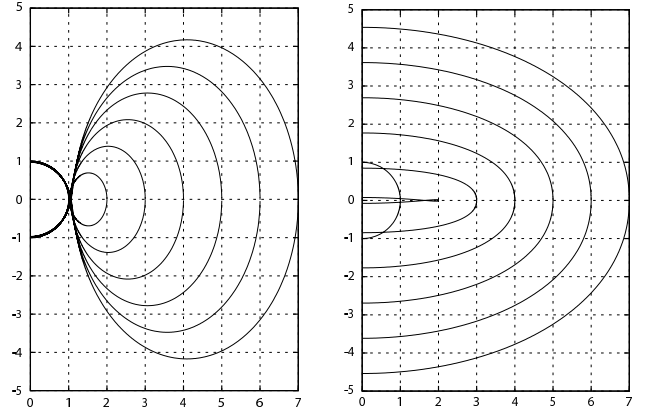}%\
\vspace{-5mm}
\end{center}
\caption{\label{figure13}%
Deformation of the ring of particles, as in Fig.~\ref{figure12}, for ${\delta=0.005}$ (left) and ${\delta=0.8}$ (right).}
\end{figure}

Finally, in Figs.~\ref{figure12} and~\ref{figure13} we visualize the deformation of the ring of test particles, initially at rest, as it evolves with time. It can be concluded that the circle (which may be considered as a ${(x,z)}$~section through a sphere) is deformed by the gravitational impulse into an axially symmetric pinched surface, elongated and expanding along the moving strings in the positive $z$-direction. Also, the particles which initially started at ${x>0}$ have ${v_x^-<0}$, while those with ${x<0}$ have ${v_x^->0}$.  This explicitly demonstrates the ``dragging'' effect in such spacetimes caused by the moving strings and the corresponding impulse. With a growing value of the parameter $\delta$, the deformation in the $z$-direction is bigger.

In the complementary case, in which the \emph{ring of static test particles is located in the} ${(x^+,y^+)}$~plane perpendicular to the string (see~Fig.~\ref{figure4}),
${z^+_i=0}$ which corresponds to ${\alpha^+=\frac{\pi}{2}}$. It thus follows from~\eqref{rlogR} that ${r=0}$, i.e., ${R=1}$. In such a case, the explicit junction conditions for positions simplify to
\begin{eqnarray}
&& x^-_i=(1-\delta)\, \frac{\cos\Phi}{\cos\big((1-\delta)\Phi\big)}\,x_i^+, \nonumber\\
&& y^-_i=(1-\delta)\, \frac{\sin\Phi}{\sin\big((1-\delta)\Phi\big)}\,y_i^+, \nonumber\\
&& z^-_i=0, \nonumber\\
&& t^-_i\,=(1-\delta) \,t_i^+, \label{interakcni polohy pro strunuz}
\end{eqnarray}
and the coefficients in Eqs.~\eqref{interakcni rychlosti string} relating the velocities on both sides of the impulse become
\begin{eqnarray}
&& d_x = -\delta\frac{1-\dpul}{1-\delta}\cos\Phi, \nonumber \\
&& d_y = -\delta\frac{1-\dpul}{1-\delta}\sin\Phi, \nonumber \\
&& d_z = 0, \nonumber \\
&& d_t = \frac{1-\delta+\dpul^2}{1-\delta} . \label{abcdStringtgenz}
\end{eqnarray}
Since ${\dot{z}_i^-=0=z_i^-}$, motion of the particles will remain in the perpendicular $(x^-,y^-)$~plane behind the impulse. In fact,
\begin{eqnarray}
&& v^-=\sqrt{(v_x^-)^2+(v_y^-)^2} = \delta \frac{1-\dpul}{1-\delta+\dpul^2}, \nonumber\\
&& \frac{v_y^-}{v_x^-} = \tan\Phi. \label{perpemotion}
\end{eqnarray}
This implies geometrically that all the particles will move radially inward with the same speed, and the circular ring will thus uniformly contract. This is in full agreement with the corresponding partial result obtained previously in Sec.~IV.~B. of \cite{PodSte03}.

For more general situations, in which the test particles in front of the impulse are not static in the ${x,y,z}$ coordinates, the resulting motion can similarly be investigated using the relations
\eqref{interakcni polohy pro strunup} and \eqref{interakcni rychlosti za vlnou pomoci pred vlnou}. In particular, employing \eqref{abcdStringxgenp}--\eqref{abcdStringtgenp} and \eqref{hyperb} for the case when ${\Phi=0}$ we obtain
\begin{eqnarray}
&& x^-_i=(1-\delta)\, x_i^+, \nonumber\\
&& y^-_i=0=y_i^+, \nonumber\\
&& z^-_i=(1-\delta)\, \frac{\sinh r}{\sinh \big((1-\delta)r\big)}\,z_i^+, \nonumber\\
%&& z^-_i=(1-\delta)\, \frac{R-R^{-1}}{R^{1-\delta}-R^{\delta-1}}\,z_i^+, \label{interakcnistrunupreal}\\
&& t^-_i\,=(1-\delta)\, \frac{\cosh r}{\cosh \big((1-\delta)r\big)}\,t_i^+, \label{interakcnistrunupreal}
\end{eqnarray}
and
\begin{eqnarray}
&& \dot{x}_i^- =  \frac{1-\delta+\dpul^2}{1-\delta}\,\dot{x}_i^+  \nonumber\\
&& \hskip6mm     -\delta\frac{1-\dpul}{1-\delta}\Big[\sinh \big((1-\delta)r\big)\,\dot{z}_i^+ +\cosh \big((1-\delta)r\big)\,\dot{t}_i^+\Big], \nonumber \\
&& \dot{y}_i^- =  \dot{y}_i^+, \nonumber \\
&& \dot{z}_i^- =  \delta\frac{1-\dpul}{1-\delta}\sinh r\,\dot{x}_i^+   \label{rychlosti za vlnou string}\\
&& \hskip6mm    +\frac{(1-\dpul)^2}{1-\delta} \Big[\cosh (\delta r)\,\dot{z}_i^+ -\sinh (\delta r)\,\dot{t}_i^+\Big], \nonumber\\
&& \hskip6mm    -\frac{\ddctvrt}{1-\delta} \Big[ \cosh \big((2-\delta)r\big)\,\dot{z}_i^+ + \sinh \big((2-\delta)r\big)\,\dot{t}_i^+ \Big], \nonumber
\end{eqnarray}
\begin{eqnarray}
&& \dot{t}_i^-\,=-\delta\frac{1-\dpul}{1-\delta} \cosh r\,\dot{x}_i^+   \nonumber\\
&& \hskip6mm    +\frac{(1-\dpul)^2}{1-\delta} \Big[ -\sinh (\delta r)\,\dot{z}_i^+ +\cosh (\delta r)\,\dot{t}_i^+\Big], \nonumber\\
&& \hskip6mm    +\frac{\ddctvrt}{1-\delta} \Big[\sinh \big((2-\delta)r\big)\,\dot{z}_i^+ +\cosh \big((2-\delta)r\big)\,\dot{t}_i^+\Big], \nonumber
\end{eqnarray}
where $r$ is given by~\eqref{rlogRy=o}.

\subsection{Effect on particles comoving in de~Sitter space}
\label{subsec:comovdeSitter}

\begin{figure}[h]
\begin{center}
\includegraphics[scale=0.7]{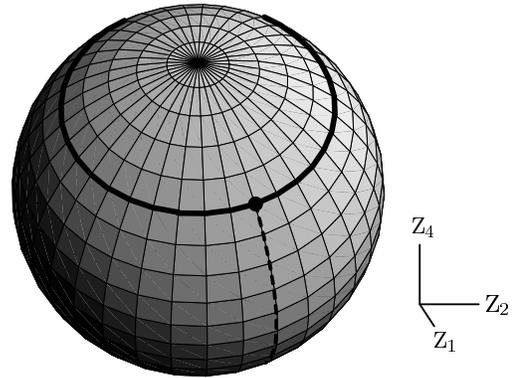}%\
\vspace{-5mm}
\end{center}
\caption{\label{figure14}%
The de~Sitter universe with the snapped cosmic string (indicated by a dashed line at ${\Z_2 = 0 = \Z_3}$) and the related impulse (${\Z_4 = a}$) at a given time
(the coordinate ${\Z_3}$ is suppressed). As the universe expands, the impulse propagates from the north pole to the equator.}
\end{figure}

Finally, it will be illustrative to investigate the effect of the impulsive spherical wave generated by a snapped cosmic string on test particles which
are comoving in the de Sitter (half-)space in front of the impulse. Specifically, these particles are initially given by
\begin{equation}
\chi=\chi^+_i=\chi_0, \quad \theta=\theta^+_i=\theta_0, \quad \phi=\phi^+_i=\phi_0, \label{comovconst}
\end{equation}
where ${\chi_0,\theta_0, \phi_0}$ are constants, in the coordinates which naturally cover the de~Sitter universe in the standard form of the metric
\begin{equation}
   \d s_0^2 = -\d t^2+a^2 \cosh^2\frac{t}{a}\left(\d \chi^2+\sin^2\chi \,(\d \theta^2+\sin^2\theta \,\d \phi^2)\right).
\label{deSitter}
\end{equation}
Such a parametrization of the de~Sitter hyperboloid~\eqref{hyperboloid} is obtained by
\begin{eqnarray}
   && \Z_0 = a\sinh\frac{t}{a} , \nonumber\\
   && \Z_1 = a \cosh\frac{t}{a}\sin\chi\cos\theta , \nonumber\\
   && \Z_2 = a \cosh\frac{t}{a}\sin\chi\sin\theta\cos\phi , \label{Zcoordsstand}\\
   && \Z_3 = a \cosh\frac{t}{a}\sin\chi\sin\theta\sin\phi , \nonumber\\
   && \Z_4 = a\cosh\frac{t}{a}\cos\chi , \nonumber
\end{eqnarray}
where ${t\in(-\infty,+\infty)}$, ${\chi,\theta\in[0,\pi]}$, ${\phi\in[0,2\pi]}$. Inversely,
\begin{eqnarray}
   && \sinh\frac{t}{a}=\frac{Z_0}{a}\,, \quad \tan^2\chi=\frac{Z_1^2+Z_2^2+Z_3^2}{Z_4^2}\,,\nonumber\\
   && \tan^2\theta=\frac{Z_2^2+Z_3^2}{Z_1^2}\,, \quad \tan\phi=\frac{Z_3}{Z_2}\,.\label{invZcoordsstand}
\end{eqnarray}
The expanding impulse is located at ${\Z_4 = a}$ (see Fig.~\ref{figure1}), i.e., it is given by ${ \cosh(t/a)=1/\cos\chi}$ which can be rewritten as
\begin{equation}
   \tanh\frac{t}{a}=\sin\chi\,.
\label{impulsedeSitter}
\end{equation}
The snapped cosmic string is located at ${\Z_2 = 0 = \Z_3}$ in the de~Sitter region in front of the impulse, which corresponds to ${\theta^+=0,\pi}$.
The spacetime can thus be visualized as in Fig.~\ref{figure14}.

Notice that the impulse is always located at the fixed value ${\,\Z_4 = a\,}$ but, as the spherical de~Sitter universe expands, the impulse propagates from
its north pole ${\chi=0}$ at ${t=0}$ to its equator ${\chi=\frac{\pi}{2}}$ as ${t\to\infty}$. The cosmic string was initially a closed loop around the
whole meridian ${\theta=0,\pi}$, but it snapped in north pole at ${t=0}$ (when the universe had the minimum radius $a$) generating the impulsive gravitational wave.

The convenient form of the junction conditions for geodesics is given in the five-dimensional representation by Eqs.~\eqref{interakcni polohyZ} and
\eqref{navazani_rychlosti_5D}. Using~\eqref{Ziexplicstruna}, \eqref{Zcoointer}, \eqref{Zcoordsstand} and \eqref{impulsedeSitter} we obtain a simple expression for the complex interaction parameter
\begin{equation}
 Z_i^{1-\delta} =\tan\frac{\theta_0}{2}\,e^{\,\im\,\phi_0}\,,
\label{Ziimpulse5D}
\end{equation}
(notice that this is consistent with the stereographic interpretation shown in Fig.~\ref{figure2}). In view of~\eqref{plarparZi} we thus obtain
\begin{equation}
r=\frac{1}{1-\delta}\log \left(\tan\frac{\theta_0}{2}\right), \qquad \Phi=\frac{\phi_0}{1-\delta}\,. \label{rlogRplus5D}
\end{equation}
In terms of these initial data we may rewrite \eqref{interakcni polohyZ} and \eqref{navazani_rychlosti_5D}, employing~\eqref{hyperb}, explicitly as
\begin{eqnarray}
&& \Z_{0i}^-=(1-\delta)\, \frac{\cosh r}{\cosh \big((1-\delta)r\big)}\,\Z_{0i}^+\,, \nonumber\\
&& \Z_{1i}^-=(1-\delta)\, \frac{\sinh r}{\sinh \big((1-\delta)r\big)}\,\Z_{1i}^+\,, \nonumber\\
&& \Z_{2i}^-=(1-\delta)\, \frac{\cos\Phi}{\cos\big((1-\delta)\Phi\big)}\,\Z_{2i}^+\,, \label{interakcni polohyZstr}\\
&& \Z_{3i}^-=(1-\delta)\, \frac{\sin\Phi}{\sin\big((1-\delta)\Phi\big)}\,\Z_{3i}^+\,, \nonumber\\
&& \Z_{4i}^- \,=\, a\,=\ \Z_{4i}^+\,.  \nonumber
\end{eqnarray}
and
\begin{equation}
\left(\begin{array}{ccccc@{\ }r}
\dot{\Z}_{2i}^{-} \\
\dot{\Z}_{3i}^{-} \\
\dot{\Z}_{1i}^{-} \\
\dot{\Z}_{0i}^{-} \\
\dot{\Z}_{4i}^{-}
\end{array}\right)
=
\left(\begin{array}{ccccc@{\ }r}
a_x & b_x & c_x & d_x & n_x \\
a_y & b_y & c_y & d_y & n_y \\
a_z & b_z & c_z & d_z & n_z \\
a_t & b_t & c_t & d_t & n_t \\
0  & 0  & 0  & 0  & 1
\end{array}\right)
\left(\begin{array}{ccccc@{\ }r}
\dot{\Z}_{2i}^{+} \\
\dot{\Z}_{3i}^{+} \\
\dot{\Z}_{1i}^{+} \\
\dot{\Z}_{0i}^{+} \\
\dot{\Z}_{4i}^{+}
\end{array}\right),
\label{navazani_rychlosti_5Dstr}
\end{equation}
where the constant coefficients ${a_\p,b_\p,c_\p,d_\p}$ are given by \eqref{abcdStringxgenp}--\eqref{hyperb} and $n_\p$ is determined by expression~\eqref{omega}.
It follows form~\eqref{Zcoordsstand} and~\eqref{impulsedeSitter} that
\begin{eqnarray}
   && \Z_{0i}^+ = a \tan\chi_0, \nonumber\\
   && \Z_{1i}^+ = a \tan\chi_0\cos\theta_0, \nonumber\\
   && \Z_{2i}^+ = a \tan\chi_0\sin\theta_0\cos\phi_0, \label{Zcoorinital}\\
   && \Z_{3i}^+ = a \tan\chi_0\sin\theta_0\sin\phi_0. \nonumber
\end{eqnarray}
Similarly, by differentiating~\eqref{Zcoordsstand} with respect to the proper time  ${\tau=t}$ of a comoving particle we obtain
\begin{eqnarray}
   && \dot{\Z}_{0i}^{+} = \frac{1}{\cos\chi_0}, \nonumber\\
   && \dot{\Z}_{1i}^{+} = \frac{\sin^2\chi_0}{\cos\chi_0}\cos\theta_0 , \nonumber\\
   && \dot{\Z}_{2i}^{+} = \frac{\sin^2\chi_0}{\cos\chi_0}\sin\theta_0\cos\phi_0 , \label{Zcoorinitalvel}\\
   && \dot{\Z}_{3i}^{+} = \frac{\sin^2\chi_0}{\cos\chi_0}\sin\theta_0\sin\phi_0 , \nonumber\\
   && \dot{\Z}_{4i}^{+} = \sin\chi_0 . \nonumber
\end{eqnarray}
These parameters explicitly satisfy the constraints \eqref{con1}, \eqref{con2}, \eqref{con3} for timelike geodesic in de~Sitter space
(${e=-1}$, ${\varepsilon=1}$)

\begin{figure}[h]
\begin{center}
\includegraphics[scale=0.38]{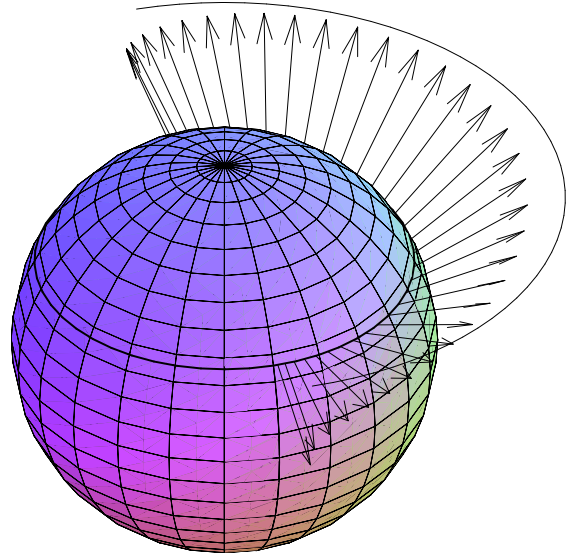}%\
\vspace{-5mm}
\end{center}
\caption{\label{figure15}%
The de~Sitter universe with the snapped string and the impulsive wave, at a given time. The arrows indicate the velocities of different test
particles behind the impulse. The outer semicircle locates the same comoving particles at a later time if the impulse would be absent, i.e., if they would move
solely due to the expansion of the universe.
}
\end{figure}

\begin{figure}[h]
\begin{center}
\includegraphics[scale=0.63]{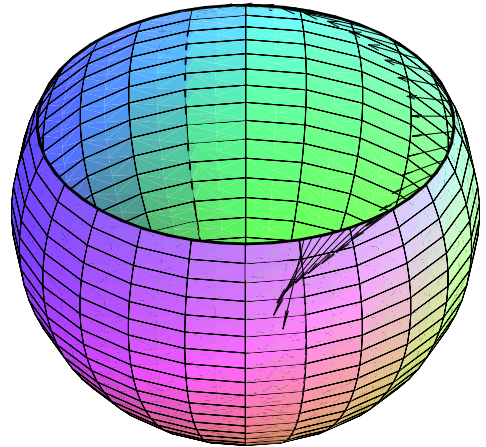}%\
\vspace{-5mm}
\end{center}
\caption{\label{figure16}%
The ``net'' effect of the impulse on the test particles, obtained by subtracting a comoving motion due to expansion of the universe. The particles are accelerated and ``dragged'' along the string.
}
\end{figure}

We can thus visualize the effect of the impulse on initially comoving particles in de~Sitter universe in the ``five-dimensional'' pictures
shown in Figs.~\ref{figure15} and~\ref{figure16}, where we plot the corresponding velocity vectors (with the spherical space, impulse and the snapped
string as in Fig.~\ref{figure14}). 

In Fig.~\ref{figure15} the arrows indicate the velocities of different test particles, given
by~\eqref{navazani_rychlosti_5Dstr} and~\eqref{Zcoorinitalvel} with the same values of $\chi_0$ (and $\phi_0$ suppressed), behind the impulsive wave.
The outer semicircle indicates the position of the same particles if the impulse would be absent --- they would (comovingly) move because the de~Sitter
universe itself expands. Therefore, the \emph{difference} gives the ``net'' effect of the impulse on these particles (by subtracting a natural comoving
motion due to the global expansion of the universe). This is shown in Fig.~\ref{figure16}. It can be seen that the particles close to the string are
accelerated to higher speeds and are ``dragged'' along the string, while the particles in the transverse plane are accelerated ``perpendicularly''
to the string. In fact, Fig.~\ref{figure7} can be understood as a projection onto the horizontal section ${\Z_4=a}$ through Fig.~\ref{figure16}.

\section{Conclusions}

We  presented a complete and explicit solution of geodesic motion which describes the effect of expanding spherical impulsive gravitational waves propagating in constant-curvature backgrounds, provided the trajectories of test particles are of class $C^1$ in a continuous coordinate system. This generalizes results obtained previously for Minkowski background space \cite{PodSte03}
to any value of the cosmological constant, i.e., the de~Sitter universe (${\Lambda>0}$) or anti-de~Sitter universe (${\Lambda<0}$). Also, it is a
counterpart of paper \cite{PodOrt01} in which motion of test particles in these background spaces with nonexpanding impulses was analyzed.

We derived a convenient form of the junction conditions~\eqref{interakcni polohy za vlnou pomoci pred vlnou}--\eqref{explicoef} and the corresponding
refraction formulae~\eqref{obecna refrakcni formule - polohy}, \eqref{obecna refrakcni formule - rychlosti}, employing the natural coordinates in
which the background metric~\eqref{mink} is conformally flat. Interestingly, the expressions are independent of the parameter ${\epsilon=-1, 0, +1}$ which occurs the 
continuous metric \eqref{en0} for the impulsive-wave spacetimes. We also considered the five-dimensional formalism  which is suitable when ${\Lambda\not=0}$, see
equations \eqref{interakcni polohyZ} and~\eqref{navazani_rychlosti_5D}.

Subsequently, we discussed in detail the behaviour of test particles in axially symmetric spacetimes in which the gravitational impulse is generated by a snapped cosmic string. In particular, we demonstrated that the particles are dominantly dragged by the impulse in the direction of the moving strings, and are accelerated to ultrarelativistic speeds in their vicinity, see Figs.~\ref{figure7}--\ref{figure9}. These results apply to any value of the cosmological constant. The strings and the associated impulse would thus effectively create opposite ``beams'' of particles, dominantly moving along the strings with the speed close to the speed of light, as visualized in Figs.~\ref{figure12} and~\ref{figure13}.

\acknowledgments

J.~P. was supported by the grant GA\v{C}R~202/08/0187 and R.~\v{S}. by the grants GA\v{C}R~205/09/H033 and GAUK~259018. 
This work was also partially supported by the Czech Ministry of Education under the project MSM0021610860.
We are grateful to Jerry Griffiths for some useful comments on the manuscript.

\end{document}